# Seafloor geodetic constraints on interplate coupling of the Nankai Trough megathrust zone

Yusuke Yokota[1], Tadashi Ishikawa[1], Shun-ichi Watanabe[1], Toshiharu Tashiro[1] and Akira Asada[2]

[1]Hydrographic and Oceanographic Department, Japan Coast Guard, 2-5-18, Aomi, Koto-ku, Tokyo 135-0064, Japan (eisei@jodc.go.jp)

[2]Institute of Industrial Science, University of Tokyo, 4-6-1 Komaba, Meguro-ku, Tokyo 153-8505, Japan (asada@iis.u-tokyo.ac.jp)

**Interplate megathrust earthquakes have inflicted catastrophic damage on human society. An extremely hazardous megathrust earthquake is predicted to occur along the Nankai Trough off southwestern Japan, an economically active and densely populated area with historical records of megathrust earthquakes[1-5]. Megathrust earthquakes are the result of a plate subduction mechanism and occur at interplate slip-deficit (or coupling) regions[6-7]. Many past studies have attempted to capture slip-deficit rate (SDR) distributions for assessing future earthquake disasters[8-10]. However, they could not capture a total view of the megathrust earthquake source region because they had no seafloor geodetic data. The Hydrographic and Oceanographic Department of the Japan Coast Guard (JHOD) has been developing a highly precise and sustainable seafloor geodetic observation network[11] deployed in this subduction zone to broadly obtain direct information related to offshore SDR. Here, we present seafloor geodetic observation data and an offshore interplate SDR distribution model. Our data suggests that most offshore regions in this subduction zone have positive SDRs. Specifically, our observations indicate previously unknown high-SDR regions that are important for tsunami disaster mitigation and low-SDR regions that are consistent with distributions of shallow slow earthquakes and subducting seamounts. This is the first direct evidence suggesting that coupling conditions are related to these seismological and geological phenomena. These observations provide new fundamental information for inferring megathrust earthquake scenarios and interpreting research on the Nankai Trough subduction zone.**

Recurring interplate megathrust earthquakes have occurred along the Nankai Trough subduction zone between the Philippine Sea plate and the Amur plate, and the next earthquake is predicted to occur in the near future[1-5]. This subduction zone is frequently discussed in terms of segmented source regions called the Nankaido, Tonankai and Tokai regions, and M8-class earthquakes in these segments are described in the past 300 years of historical records[2] (the 1707 Hoei, 1854 Ansei-I, Ansei-II, 1944 Tonankai and 1946 Nankaido earthquakes). Up to M9-class earthquakes are believed to have occurred along each segment[5].

Since megathrust earthquakes are driven by accumulated interplate slip-deficit, these historical earthquakes were believed to occur on an interplate boundary with a high SDR[6-7]. To assess the scale of future earthquakes and tsunamis, it is therefore necessary to grasp the whole interplate SDR distribution. While many geodetic approaches have been attempted to obtain this information for the Nankai Trough, they have not been successful. This is because the previous geodetic observation network is biased to land areas and cannot capture total geodetic information on the seafloor above the interplate boundary[8-10]. Although small-scale seafloor geodetic observations have been carried out[12], observations were limited to only around the Kumano-nada region.

Accordingly, over the past decade we have taken a new approach to obtaining total seafloor geodetic information by means of a broad-scale seafloor observation network using the Global Positioning System and Acoustic ranging combination (GPS-A) technique[11,13-14]. The precision and frequency of our GPS-A observations have been improved by original developments since 2000 and are among the highest standards in the world. Our GPS-A technique is described in the Methods section and Extended Data Fig. 1.

We observe fifteen seafloor sites in a wide seafloor region along the Nankai Trough (Fig. 1). Six sites were established before the 2011 Tohoku-oki earthquake. Since the results of these sites were insufficient to expose a total image of interplate SDR[15], nine new sites were deployed after 2011.

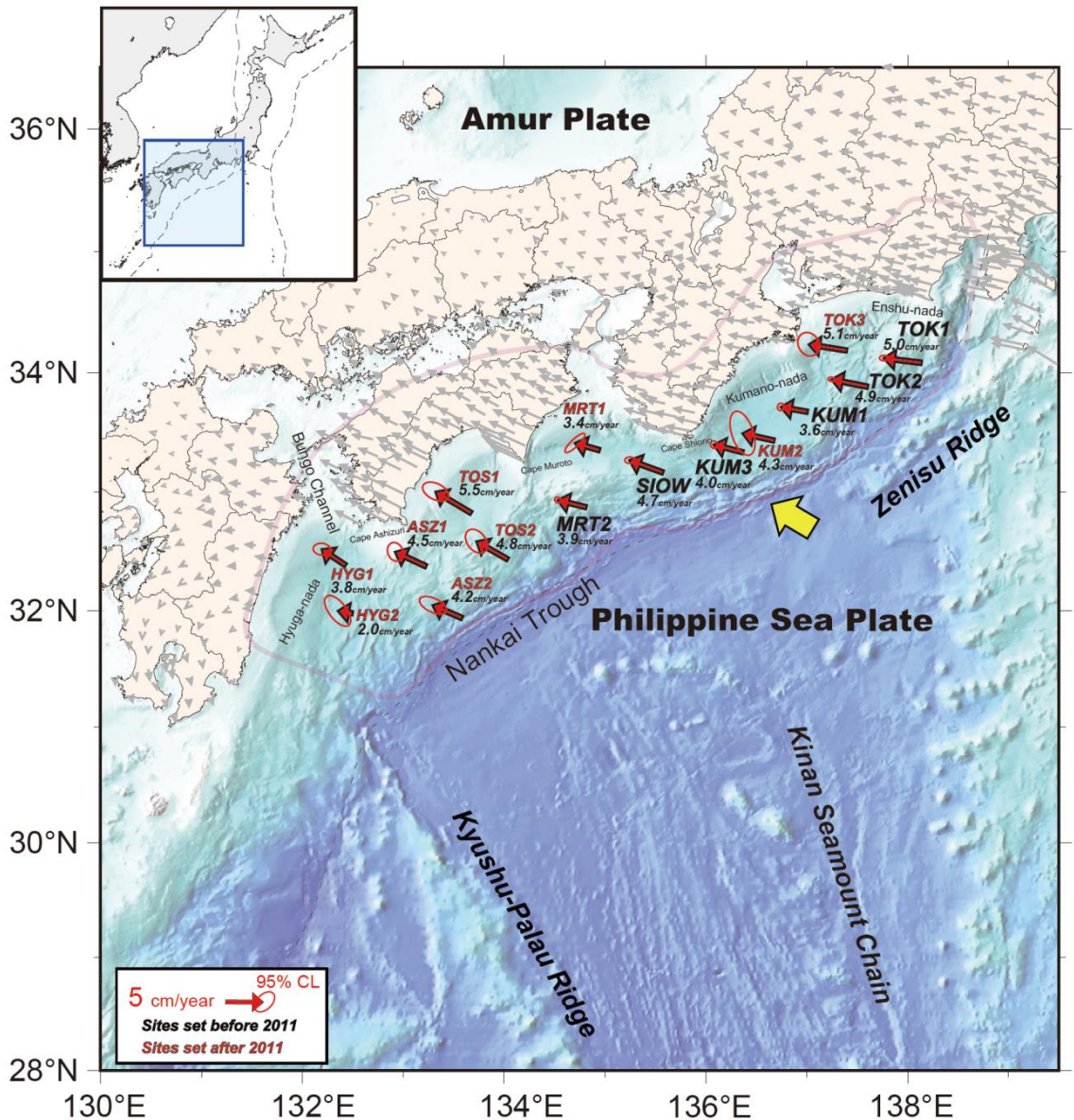

**Fig. 1 | Seafloor velocity field from seafloor geodetic observations at fifteen seafloor sites along the Nankai Trough.**

Seafloor velocity vectors are shown by red arrows. Each ellipse indicates the 95 % confidence level. The onshore velocity vectors are calculated for the period from March 2006 to December 2009 using GEONET stations and shown by light-gray arrows. Yellow arrow indicates the convergence rate (6.5 cm/year) of the Philippine Sea plate under the Amur plate calculated using the MORVEL model[17]. Purple region shows the maximum source region provided as the worst case scenario by the Central Disaster Management Council of the Japanese Government[5]. Seafloor topography was based on J-EGG500 of the Japan Oceanographic Data Center (JODC) of the JHOD.

(1) TOK1

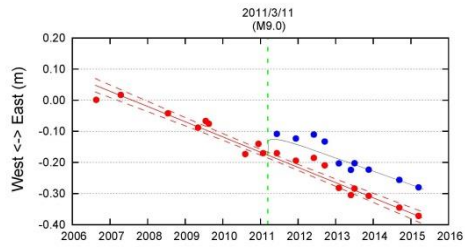
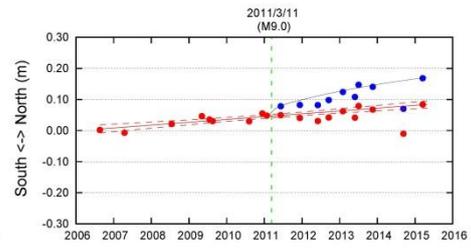

(2) TOK2

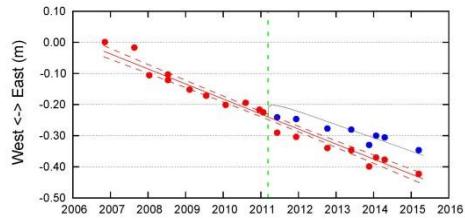
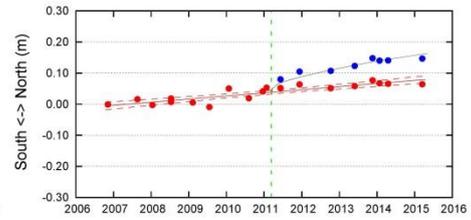

(3) TOK3

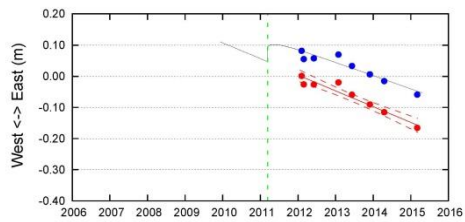
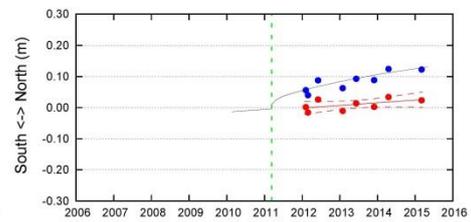

(4) KUM1

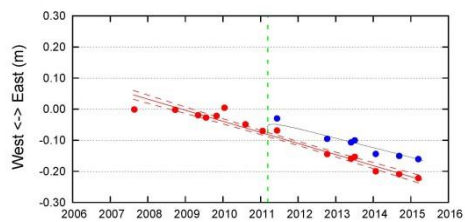
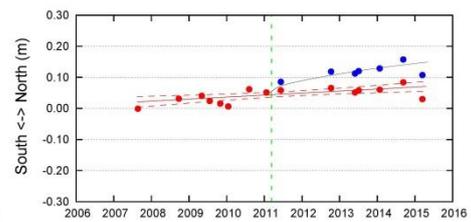

(5) KUM2

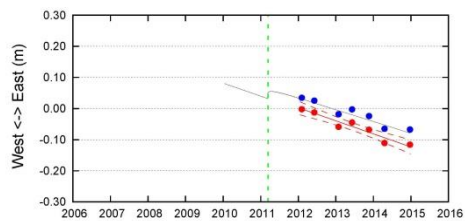
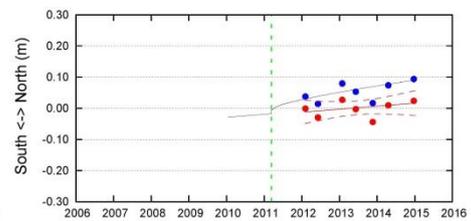

(6) KUM3

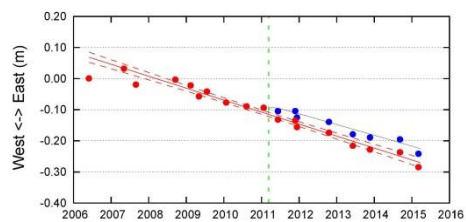
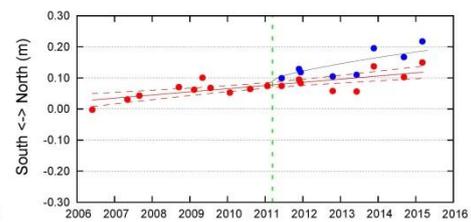

(7) SIOW
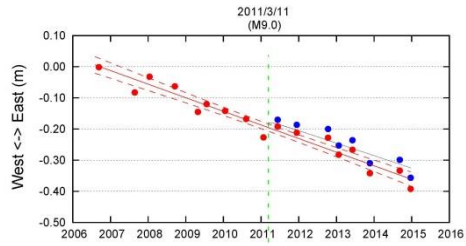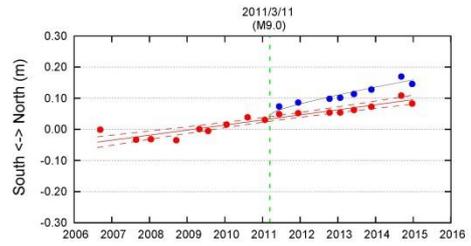

(8) MRT1
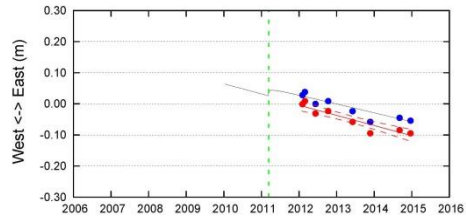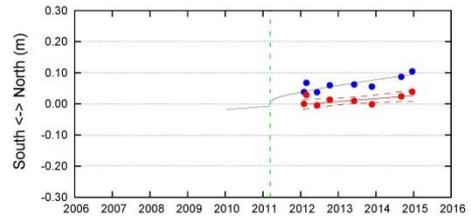

(9) MRT2
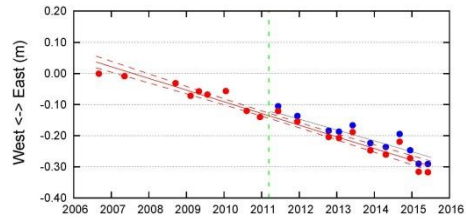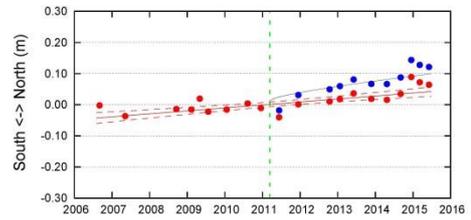

(10) TOS1
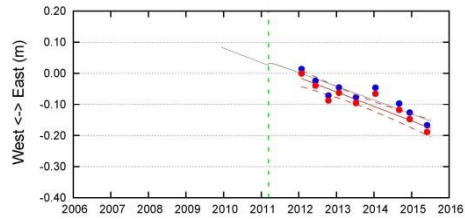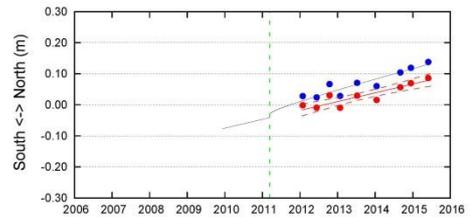

(11) TOS2
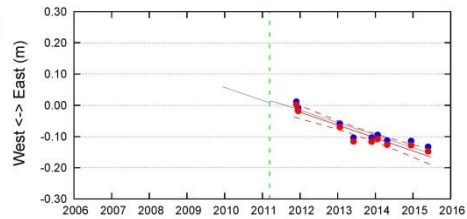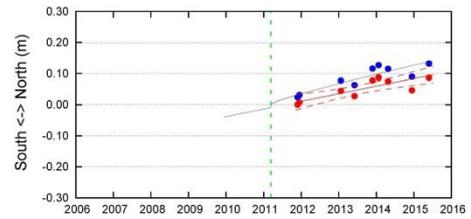

(12) ASZ1
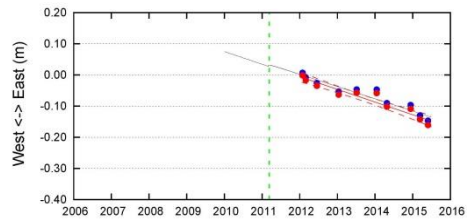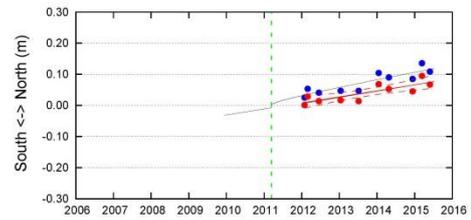

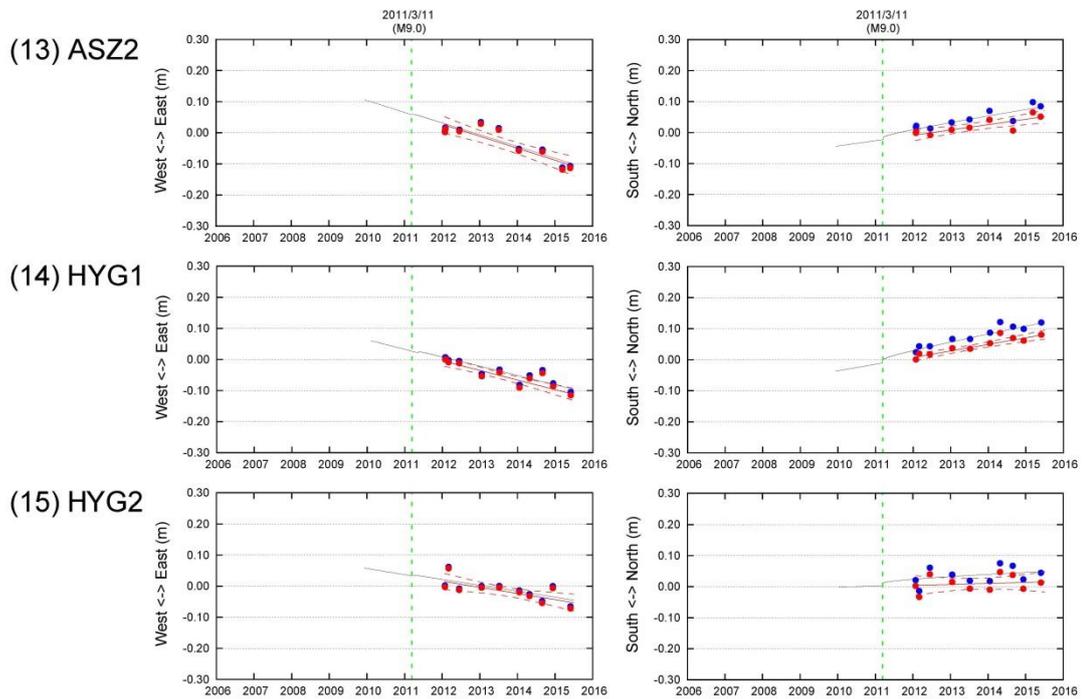

**Extended Data Figures 2-4 | Time series in east-west (left column) and north-south (right column) displacements.** These data were obtained at the fifteen seafloor sites for the period after 2006. The position reference is the Amur plate. Blue circles indicate the raw observations before deductions of the coseismic and postseismic effects due to the 2011 Tohoku-oki earthquake. Red circles indicate the corrected final results. The linear trends and the 95% two-sided confidence intervals are shown with red solid and dashed lines, respectively. Black lines show the calculated coseismic and postseismic deformations following the Tohoku-oki earthquake[18-19].

| Site name | Position | | Velocity (cm/year) | | | Standard deviation and Correlation (cm/year) | | |
|---|---|---|---|---|---|---|---|---|
| | latitude | longitude | ABS ($V$) | east ($V_e$) | north ($V_n$) | $\sigma(e)$ | $\sigma(n)$ | $Corr(e, n)$ |
| TOK1 | 34.08 | 138.14 | 5.0 | -4.9 | 0.9 | 0.2 | 0.1 | 0.0 |
| TOK2 | 33.88 | 137.61 | 4.9 | -4.8 | 1.0 | 0.2 | 0.1 | -0.1 |
| TOK3 | 34.18 | 137.39 | 5.1 | -5.1 | 0.8 | 0.4 | 0.5 | -0.1 |
| KUM1 | 33.67 | 137.00 | 3.6 | -3.6 | 0.7 | 0.1 | 0.2 | 0.1 |
| KUM2 | 33.43 | 136.67 | 4.3 | -4.2 | 1.0 | 0.5 | 0.9 | -0.5 |
| KUM3 | 33.33 | 136.36 | 4.0 | -3.9 | 1.0 | 0.2 | 0.2 | -0.1 |
| SIOW | 33.16 | 135.57 | 4.7 | -4.4 | 1.6 | 0.2 | 0.2 | 0.0 |
| MRT1 | 33.35 | 134.94 | 3.4 | -3.3 | 1.0 | 0.4 | 0.4 | 0.9 |
| MRT2 | 32.87 | 134.81 | 3.9 | -3.8 | 1.0 | 0.2 | 0.2 | -0.1 |
| TOS1 | 32.82 | 133.67 | 5.5 | -4.7 | 2.8 | 0.6 | 0.4 | -0.4 |
| TOS2 | 32.43 | 134.03 | 4.8 | -4.2 | 2.4 | 0.5 | 0.5 | -0.4 |
| ASZ1 | 32.37 | 133.22 | 4.5 | -4.1 | 1.9 | 0.3 | 0.4 | -0.2 |
| ASZ2 | 31.93 | 133.58 | 4.2 | -3.9 | 1.7 | 0.6 | 0.4 | -0.5 |
| HYG1 | 32.38 | 132.42 | 3.8 | -3.1 | 2.1 | 0.4 | 0.3 | -0.1 |
| HYG2 | 31.97 | 132.49 | 2.0 | -2.0 | 0.3 | 0.6 | 0.7 | -0.6 |

**Table 1 | Velocity of each site with respect to the Amur plate with standard deviations and correlation.**

Extended Data Figs. 2, 3 and 4 show time series of the estimated horizontal coordinates of seafloor sites for each epoch relative to their locations in the first campaign. The reference frame is ITRF2005[16]. The positions are presented with respect to the stable part of the Amur plate, based on the MORVEL velocity model[17]. The position data of each epoch is summarized in Supplementary Table 1. Raw data from the six original sites involved coseismic deformation steps due to the Tohoku-oki earthquake. We removed these steps from raw data based on a coseismic source model established using onshore and seafloor geodetic data[18]. In addition, raw data from all sites involved postseismic deformations due to afterslip and viscoelastic relaxation following the Tohoku-oki earthquake. This nonlinear deformation was removed using the postseismic model calculated by means of a 3-D finite element method[19].

The corrected data gave us seafloor velocity fields that reflect strain accumulation processes due to the Philippine Sea plate subduction under the upper Amur plate when all sites moved at stable displacement rates. Extended Data Figs. 2, 3 and 4 present linear trends fitted to the time series using a robust regression method (the $M$ estimation method). Table 1 lists each velocity. The lines fitted to the EW and NS position time series in Extended Data Figs. 2, 3 and 4 represent the estimated linear station velocities and are shown with their 95% confidence intervals, which are used for the confidence ellipses shown in Fig. 1. The vertical velocities were not detected because they were smaller than detection limit (3 - 4 cm/year).

Long observation periods of the six original sites made the confidence ellipses small. These velocity fields are also compared with onshore GNSS data calculated for the stable period from March 2006 to December 2009 and the convergence rates of the Philippine Sea plate to the Amur plate in Fig. 1 calculated using the MORVEL model. These seafloor data are roughly consistent with the orientation of plate convergence and onshore velocities. Therefore, all offshore regions on the interplate boundary have positive SDRs.

These new data have great potential for advancing SDR distribution estimation[20]. Onshore data has no resolving power for offshore interplate boundaries, as described in the Methods section. Our seafloor data can show the offshore heterogeneity, though regions adjacent to the trench axis other than near TOK1 and ASZ2 cannot be resolved. The SDR distribution model established using these seafloor geodetic data is shown in Fig. 2a. The inversion strategy and detailed information are described in the Methods section, Supplementary Tables 2 and 3.

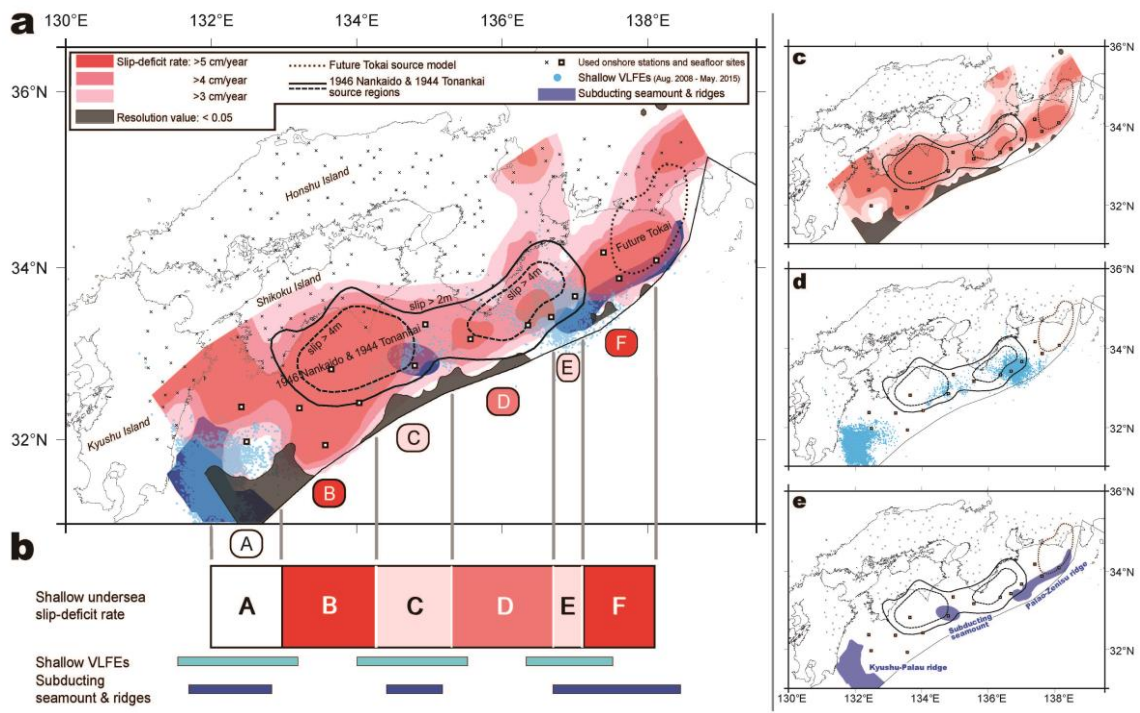

**Fig. 2 | Interplate SDR distribution indicated by the onshore and seafloor geodetic data.**

**a.** Contour map indicates the SDR distribution (SDR: > 3 cm/year) obtained using the onshore and seafloor geodetic data. Light-blue dots and blue polygons indicate the shallow VLFEs[24] and subducting seamount and ridges[21-23], respectively. Dotted and solid polygons show the future Tokai source model[5] and the assumed large slip (> 2 m) regions due to the 1946 Nankaido and 1944 Tonankai earthquakes[2], respectively. Gray shadings indicate areas with resolution values lower than 0.05. **b.** A schematic illustration of the segmented source regions along the Nankai Trough. Upper figure shows a total conceptual illustration of the segmented high-SDR regions in the shallow side of the Nankai Trough. Dark colored region indicates a high-SDR. The regions of shallow VLFEs and subducting seamounts are shown as bottom light-blue and blue bars, respectively. **c, d and e.** Separate figures of SDR distribution, shallow VLFEs and subducting seamounts.

Along the Nankai Trough, subducting seamounts are located in three regions[21-23] where the very-low-frequency earthquakes (VLFEs) were activated[24]. We mainly discuss the relation of the shallow SDR distribution with these seismological and geological features and the latest and predicted megathrust earthquake source regions from west to east. The deep part of this model is robustly similar to those in past studies using only onshore data[8-10].

In region-A, an edge of the model region cannot be resolved enough. However, only our westernmost site HYG2 directly could catch a glimpse of the undersea SDR. The displacement rate at HYG2 was lower than those at adjacent sites (HYG1, ASZ1 and ASZ2) with certain confidence levels (95 %, 95 % and 90 % CL) according to the parallelism tests between each EW component. These data and our model suggest the VLFE occurrence region extending to the east of the Kyushu-Palau ridge had a lower SDR than adjacent undersea regions. This spatial relation suggests that the subducting ridge not only activates shallow VLFEs, but also forms the low-SDR region, that is, low-coupling condition.

In region-B, the deep high-SDR region corresponds with the latest megathrust earthquake source region. The region extends to the shallow side near the trench axis, which had no slip in the latest event. No conspicuous activity of the VLFEs or subducting seamount was detected in this near-trench region. In this high-SDR region, patches of overshot SDR (more than convergence rate: approximately 6.5 cm/year) exist as in past studies[8-9]. These are probably due to interseismic viscoelastic effects[7] or underestimation of the convergence rate.

This broad high-SDR region is segmented on the eastern region. This region-C is estimated to have a lower SDR than neighboring regions-B and D. Additionally, the VLFE activity and the subducting seamount are located together, as in region-A. This spatial correspondence is additional evidence that the three phenomena have a physical correlation.

In region-D, where the Kii Peninsula protrudes to the south, the obtained SDR distribution corresponds with the latest earthquake source regions. The high-SDR region-F also corresponds with the future Tokai source model[5]. However, it reaches to the southwest region, which had no slip in the 1944 Tonankai earthquake and was not indicated as a future Tokai source model. These regions are partitioned by low-SDR region-E.

Below the shallow seafloor from regions-D to F, the Paleo-Zenisu ridge is subducting in the region nearest to the trench axis. On the other hand, intensive VLFE activity is located in the gap region-E. Therefore, the low-SDR region-E has correlation with the VLFE activity rather than the subducting ridge, though the resolving power is insufficient in the shallower south region as compared to our seafloor sites.

Observation studies[25-26] in the subduction zone worldwide had inferred the relation of the low-SDR condition with the subducting areas in front of topographic features, including seamounts. VLFE activity was also predicted to be related with the low-SDR condition[27]. In the Nankai Trough, indirect seismological evidence inferred the physical relation of the low-SDR condition with ridges and VLFEs in recent seafloor research studies[23,28], probably due to elevated pore-fluid pressure and a complicated fracture network. Three low-SDR regions-A, C and E discovered by our observation were the first direct evidences suggesting that subducting seamounts generate VLFE activity, which has low-SDR condition. It also suggests a possibility that VLFEs are activated in the low-SDR region in front of subducting seamounts worldwide.

Fig. 2b shows a conceptual illustration of the segmented source regions discovered by our observation. This 'shallow' segmentation is inconsistent with the well-known 'deep' segmentation[2] of the Nankai Trough source region. Because the shallow high-SDR patches control the scale of tsunamis from megathrust ruptures, they are important for the assessment and early warning of the tsunami disaster. For example, the Tohoku-oki earthquake had a large amount of very shallow slip[29], which for led to a devastating tsunami. The high-SDR regions-B and F are located on the outer shallow regions of the most recent and anticipated earthquake source regions and have no historical earthquake record since the 1854 Ansei earthquakes[2]. These shallow regions have accumulated slip-deficit at least since 1854 and have the potential to drive shallow rupture and tsunami.

The low SDRs are indicated in regions-A, C and E, which segment the high-SDR (megathrust earthquake source) regions. For example, the 1944 Tonankai earthquake occurred on region-D and stopped at region-E in front of region-F[1,3-4]. However, when a rupture breaks through a segment boundary, a larger event is possible. The 1946 Nankaido earthquake progressed from region-D through region-C and finally reached the deep side of region-B[1,3-4]. In this way, the low-SDR regions control megathrust earthquake scenarios.

Our observation results and SDR distribution model reflect crustal deformation over only the last several years. We plan to perform continuous observations over decades to investigate the stability of interplate SDR distributions. We can also determine whether decadal changes in crustal deformation like those observed in eastern Japan[29] occur in this subduction zone.

**Acknowledgements**

We thank Professor R. Burgmann and Professor J.-P. Avouac for reviews and constructive comments that have helped improve this manuscript. We thank Dr. O. L. Colombo of the NASA Goddard Space Flight Center for providing us with the kinematic GPS software "IT" (Interferometric Translocation)[30] and the Geospatial Information Authority of Japan (GSI) for providing us with the high-rate GPS data for the kinematic GPS analysis and the daily coordinates of the sites on the GSI's website. The coseismic and postseismic deformations of the 2011 Tohoku-oki earthquake were calculated by Dr. T. Iinuma and T. Sun, respectively. Comments from Professor K. Wang and Associate Professor A. Kato improved our manuscript. We thank T. Yabuki of the JHOD for providing us with the Bayesian inversion software. Additionally, many among the staff of the JHOD, including the crew of the S/Vs Takuyo, Shoyo, Meiyo and Kaiyo, have supported our observations and data processing. Some figures were produced using the GMT software by P. Wessel and W. H. F. Smith.

**Author Contributions**

Y.Y. carried out the inversion analysis. T.I. designed the study and performed the statistical processing. Y.Y. and S.W. performed the resolution tests. Y.Y., T.I., S.W. and T.T. processed the GPS-A seafloor geodetic data. A.A. constructed the GPS-A system.

**Methods**

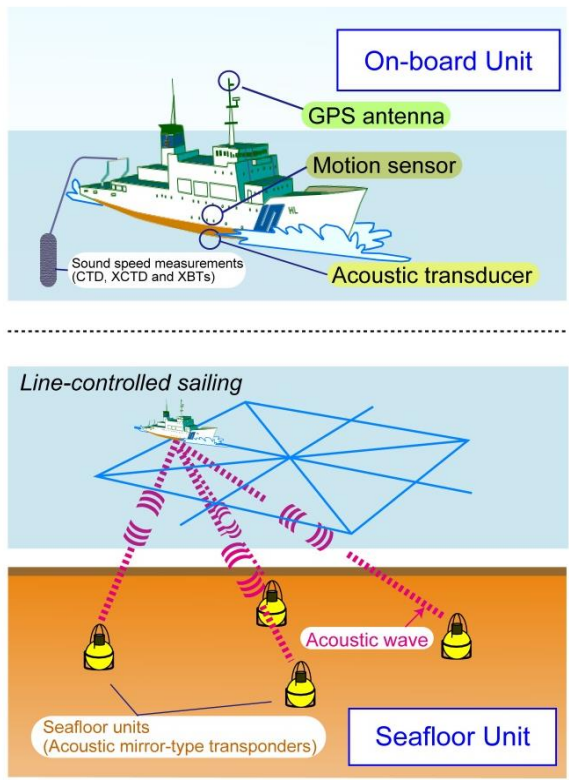

**Extended Data Figure 1 | A schematic picture of our GPS-A seafloor geodetic observation system developed by ref. 11 and improved after ref. 35.** This picture was modified from refs. 15, 34 and 37.

**Seafloor geodetic observation**

Since the radio wave scatters in the seawater, we measure the seafloor movements using the GPS observation above the sea and the acoustic ranging under the sea. This method is called as the GPS-A, which is a unique approach to monitor an absolute horizontal movement directly above the offshore interplate boundary. This technique was proposed in 1980s[31] and established after 1990s[11,13-14]. However, since their observation precision was lower than the present method, they achieved a robust observation result and needed an uneconomical very-long observation period. After 2000, the JHOD have been developing high-precise and sustainable observation techniques and provided valuable data for geodesy and seismology, e.g., the pre-, co- and post-seismic seafloor deformations of the 2011 Tohoku-oki earthquake[32-34].

A schematic picture of our present seafloor geodetic observation system[11,35-37] is shown in Extended Data Fig. 1. This system consists of a seafloor unit with four acoustic mirror-type transponders and an on-board unit with an undersea on-board acoustic transducer, a GPS antenna/receiver and a dynamic motion sensor. The on-board acoustic transducers were mounted at the stern of survey vessels for a drifting survey before 2007. After 2008, we provided a hull-mounted system to perform a line-controlled sailing survey[37] for stability and efficiency.

This system acquires three kinds of data. Kinematic GPS data are gathered to determine the absolute position of the survey vessel. Attitude data on the survey vessel are also obtained on board by a dynamic motion sensor to determine the coordinates of the on-board transducer relative to those of the GPS antenna. Distance data from the on-board transducer to the seafloor acoustic transponders are measured by acoustic ranging technique. The obtained roundtrip acoustic travel times are transformed to the ranges using sound speed profiles in seawater. These profiles are obtained using temperature and salinity profilers (conductivity temperature depth profiler (CTD), expendable conductivity temperature depth profiler (XCTD) and expendable bathythermographs (XBTs)) every several hours.

The consecutive absolute positions of the on-board transducer were determined by kinematic GPS analysis using IT software[30] and attitude data on the survey vessel. The position references are the onshore GEONET stations conducted by GSI[38]. The resultant position of the seafloor transponder was determined using a linearized inversion method based on a least squares formulation combining the absolute on-board transducer positions and the ranges to the seafloor acoustic transponders. This final analysis was constrained by the positional relationship of the grouped transponders for all epochs[36]. Here, this analysis cannot provide substantive information for positioning error of each epoch because we combine the independent observations to estimate all the positions.

To stabilize the estimates, we acquire acoustic ranging data of 3000 ~ 5000 shots for one observation at each site. We spent approximately 24 h performing an observation. Observation uncertainty of this technique is up to 2 - 3 centimeters in the horizontal component in each epoch. On the other hand, the vertical component has much uncertainty, because we observe the seafloor only from the upper region as similar as GNSS. A detection limit of the vertical velocity is 3 - 4 cm/year.

**Data processing**

Raw seafloor geodetic observation data referred to the Amur Plate in MORVEL[17] are indicated as red circles before the Tohoku-oki earthquake and blue circles after this earthquake in Extended Data Figs. 2, 3 and 4. We deducted coseismic and non-linear postseismic effects due to the Tohoku-oki earthquake from these raw data. The coseismic effects were calculated based on the coseismic source model[18] established using many onshore stations of the GNSS network (GEONET, the University of Tohoku and others), seafloor geodetic data[33,39] and data of ocean bottom pressure gauges installed by the Earthquake Research Institute, the University of Tokyo. The postseismic effects were calculated based on the deformation model[19] established using this coseismic slip model and coordinated to match the GSI's onshore data and seafloor data (observed by the Tohoku University and us[34]) following the Tohoku-oki earthquake by means of the 3-D spherical-Earth finite element model. The prototype of this viscoelastic model was established in ref. 40 by including not only the mantle wedge and the oceanic mantle but also lithosphere-asthenosphere boundary based on the afterslip model developed by ref. 29. The revised model obtained in ref. 19 with regard to shallow afterslip was used in this study. These coseismic and postseismic effects were shown in Extended Data Figs. 2, 3 and 4. These are large in the eastern sites near the source region of the Tohoku-oki earthquake and very small in the western sites. The deducted data are indicated as red circles in Extended Data Figs. 2, 3 and 4. These final data and the raw data are listed in Supplementary Table 1.

For estimations of the site velocities, we used a robust regression method ($M$ estimation method) employing Turkey's biweight function. The negative influences of outliers mainly due to disturbances of an undersea sound speed structure were mitigated.

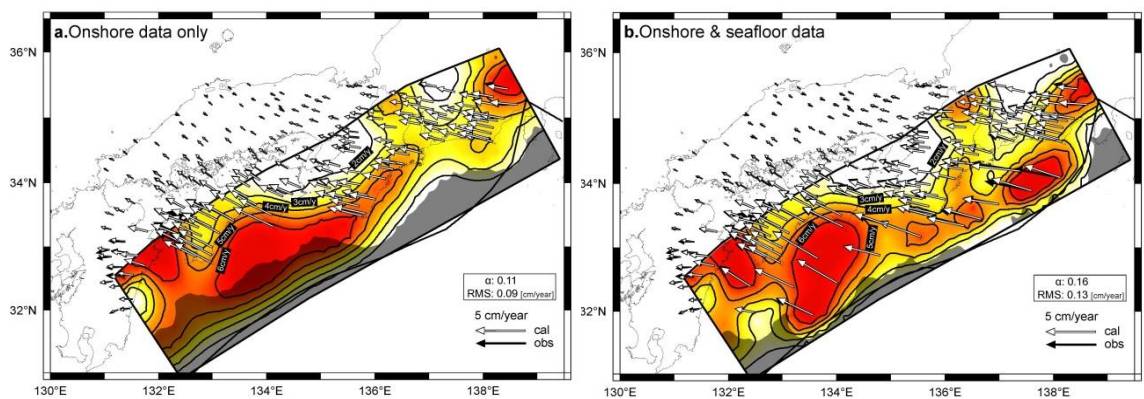

**Extended Data Figure 5 | Comparison of the calculated SDR distributions.** Resultant SDR distributions (SDR: > 2 cm/year) calculated (**a**) using only onshore data and (**b**) using onshore and seafloor data. Black and white vectors indicate the used data and calculated velocities, respectively. Gray shadings indicate areas with resolution values lower than 0.05 calculated in Extended Data Figs. 6d and 6e.

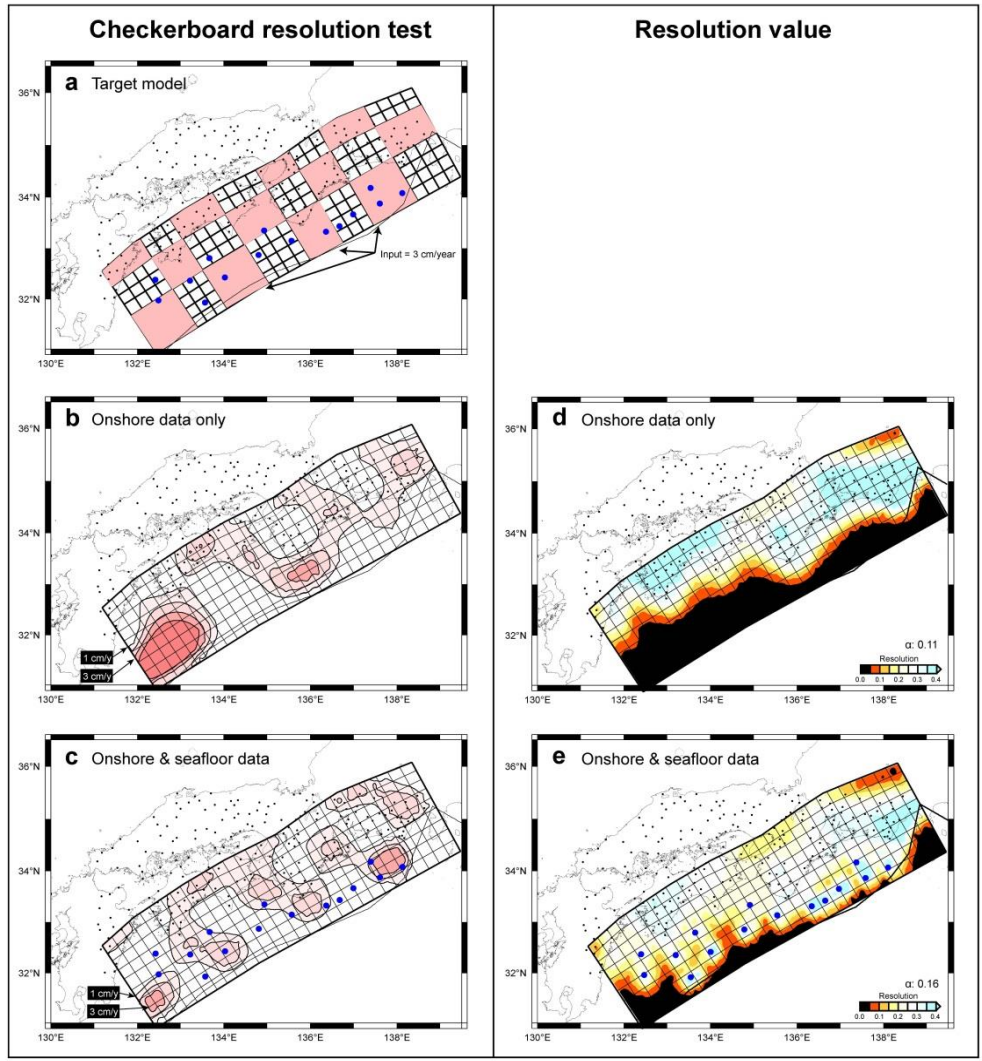

**Extended Data Figure 6 | Checker-board resolution tests and distributions of resolution values.** (Left) Results of checker-board resolution tests for the SDR inversions: **a.** Input checker-board like SDR distribution. The results of the cases **b.** using only onshore data and **c.** using both onshore and seafloor data. (Right) Resolution values as diagonal elements of the resolution matrix were also calculated for the cases **d.** using only onshore data and **e.** onshore and seafloor data, respectively.

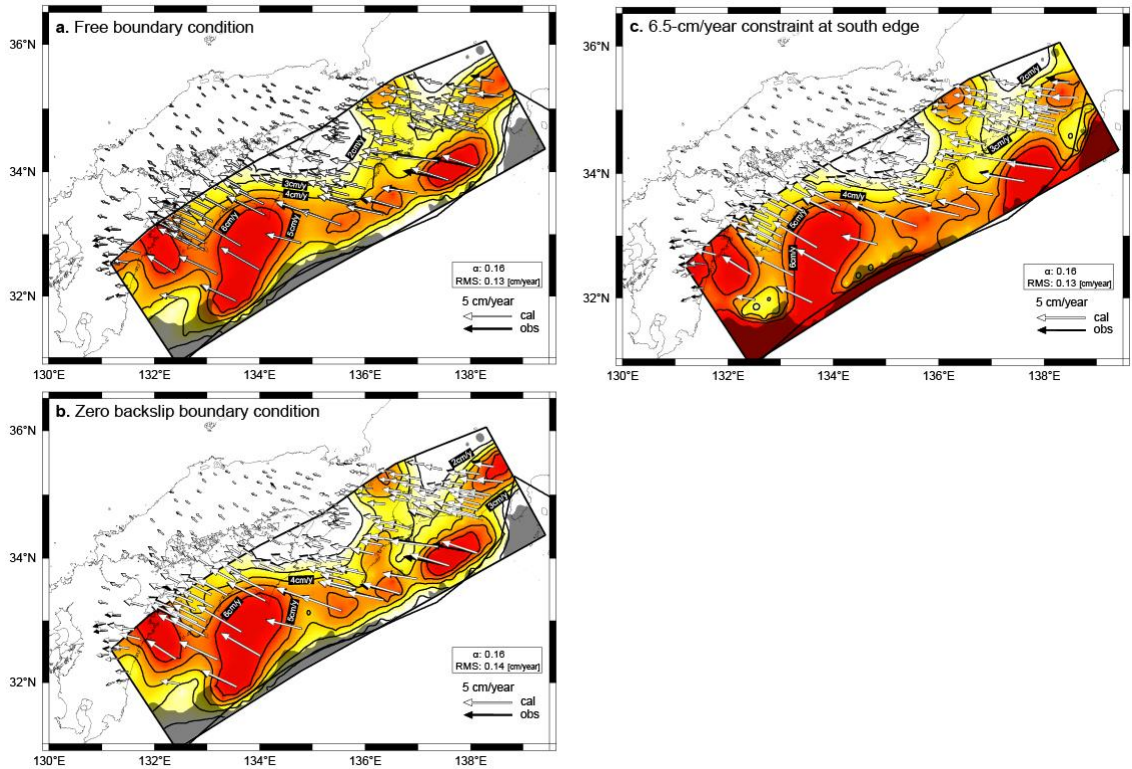

**Extended Data Figure 7 | Examinations of boundary condition effects.** Estimated SDR distributions (SDR: > 2 cm/year) calculated with (**a**) free condition, (**b**) zero backslip condition at all the boundaries and (**c**) trench high-SDR condition. Gray shadings indicate areas with resolution values lower than 0.05 calculated in Extended Data Fig. 6e.

**Interplate SDR inversion method**

We constructed the SDR distribution model by means of a geodetic inversion using the Yabuki and Matsu'ura method[20]. In this method, two prior constraints (α and σ) were included. We determined the best estimates of these hyperparameters by minimizing Akaike Bayesian Information Criterion (ABIC)[41] and obtained an optimal model. The SDR distribution model was coordinated to match our seafloor geodetic data and onshore GNSS data.

We set a fault model with approximately 800 km in the strike direction of 237° and approximately 300 km in the dip direction on the plate boundary. Our fault model was deployed simply on the interplate boundary model called as CAMP Standard Model[42]. We used a B-spline function as a basis function and calculated the SDR values by distributing subfaults on the plate boundary. Green's functions were calculated using the formulation of ref. 20 considering a homogeneous elastic half-space. We deployed the broader model than the along-strike region around our seafloor sites. Thus, we also used the onshore GNSS data of GEONET around our fault model. These onshore data were calculated for the stable period from March 2006 to December 2009. In order to avoid biases in the inversion and keep smooth resolution of the SDR model, we sub-sampled the GEONET stations. The weight functions were set equally in all the onshore and offshore stations. Model boundary conditions are detailed in subsection: Model boundary condition effect.

The northern edge of our model would be affected by the block motions, because the block boundaries around the western Japan are located near the northern boundary of our model region. However, detailed investigation of block motions[43] showed that the maximum deformation rate of the block boundary (Median Tectonic Line) does not so much affect the undersea SDR distribution (less than 8 mm/year in the northernmost region of Shikoku Island and 3 mm/year in the eastern region). Intra-plate deformation has a negligible in this SDR model calculation. Although splay faults[22,44-45] have also an implication for the megathrust earthquakes and the tsunami generation, these smaller-scale fault geometries cannot be monitored and discussed by our present seafloor geodetic observation network.

Our resultant SDR distribution model is shown in Fig. 2 and Extended Data Fig. 5b. Hyperparameter values for the prior constraints ($\alpha$ and $\sigma$) were $1.6 \times 10^{-1}$ and $1.3 \times 10^{-1}$. Our data improved the past model using only onshore data as shown in Extended Data Fig. 5a. The calculated SDR values for subfaults and the comparison between observed and calculated data are described in Supplementary Tables 2 and 3, respectively.

**Resolution of the SDR inversion**

Checkerboard resolution tests were performed for examination of these data. We generated synthetic data for the checkerboard-like SDR distributions (Extended Data Fig. 6a) with errors (2-sigma: 0.3 cm/year and 1.5 cm/year for onshore and seafloor data). The synthetic data were inverted using the same parameters and settings as the SDR inversion. Extended Data Figs. 6b and 6c indicate the resultant distributions using only onshore data and using both onshore and seafloor data, respectively. An unsolved offshore region in Extended Data Fig. 6b was solved clearly in Extended Data Fig. 6c.

Extended Data Figs. 6d and 6e also show resolution values as diagonal elements of the resolution matrix calculated for the cases using only onshore data and onshore and seafloor data, respectively. The resolution matrix was represented as follows:

$$\mathbf{R} = (\mathbf{H}^\mathrm{T}\mathbf{H} + \alpha^2 \mathbf{G}^\mathrm{T}\mathbf{G})^{-1} \mathbf{H}^\mathrm{T}\mathbf{H}$$

**H**: Static response function matrix

α: Hyperparameter of smoothness[20]

**G**: Spatial smoothness matrix[20].

Undersea areas with low values shown in Extended Data Fig. 6d were improved by the seafloor data (Extended Data Fig. 6e), though the region adjacent to the trench axis cannot be resolved even with our seafloor network because there is no site.

Additionally, Green's functions for outer subfaults of interplate boundary (on the south of trench) were set to zero. These subfaults affect neighbor low-resolution subfaults through the spatial smoothing. By these reasons, a resolving power for the shallowest subfaults to the south of our seafloor sites was not sufficient as shown in Extended Data Fig. 6c.

**Model boundary condition effect**

We examined the boundary condition of this inversion model. Each test was also calculated using the best hyperparameters determined by minimizing ABIC.

Our resultant SDR model was calculated with "zero backslip (full creeping)" condition at the trench side boundary and free condition at other model boundaries. Here, we calculated with free condition at all the boundaries, with zero backslip condition at all the boundaries and with 6.5 cm/year constraint at south edge (others: free condition), respectively, in order to examine the model boundary condition effect. Extended Data Figs. 7a, 7b and 7c showed the free condition, the zero backslip condition and the trench high-SDR condition results, respectively. These results suggest that the boundary condition did not control the main part of the undersea SDR calculation except for low-resolution shallow areas. There were small differences in the RMS of misfits between the observations and calculations in these cases.

**VLFE distribution**

The VLFE distribution used in Figs. 2a and 2d was determined by automatic analysis[24] using the method of ref. 46. This approach separates VLFEs and ordinary earthquakes automatically by comparing to NIED Hi-net catalogue. However, aftershocks following the M7 class event could not be fully differentiated. Then, we plotted the unordinary events (mainly VLFEs) in the period of August 1, 2008 – May 10, 2015 without aftershocks of the 2004 off the Kii Peninsula earthquake. We also plotted all the events detected in the periods of June 1, 2003 – May 10, 2015 (Extended Data Fig. 8).

**Subducting seamounts**

Reflection and refraction surveys were performed broadly along the Nankai Trough based on the past geomagnetic studies and seismic and bathymetric prior information. These surveys[21-23] detected three subducting seamounts (Figs. 2a and 2e) in front of the visible bathymetric features (Kyushu-Palau ridge, Kinan seamount chain and Zenisu ridge) shown in Fig. 1. The VLFEs were activated around these regions.

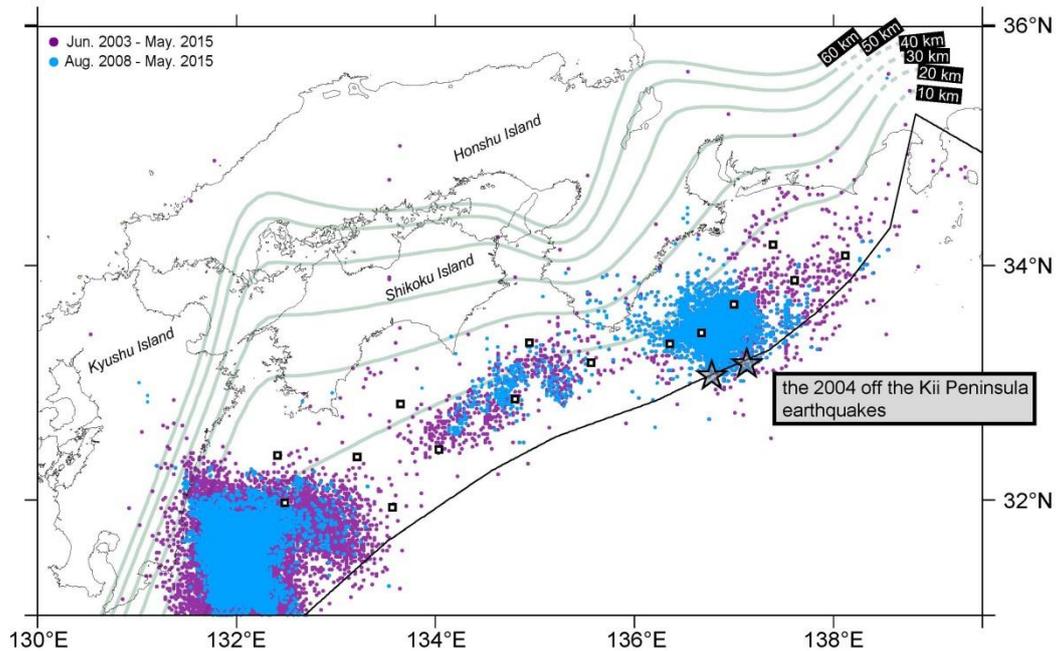

**Extended Data Figure 8 | Events determined using automatic VLFE detection analysis[24].** Purple and light-blue dots denote events in the periods of June 1, 2003 – May 10, 2015 and August 1, 2008 – May 10, 2015, respectively. Gray stars indicate epicenters of the 2004 off the Kii Peninsula earthquakes. The depths of the plate boundary of CAMP model[42] are indicated by green lines.

**Supplementary Table legends**

**Supplementary Table 1 | Estimated relative site positions.** The reference frame is ITRF2005. Eastward and Northward components indicate the data corrected the coseismic and postseismic effects due to the Tohoku-oki earthquake. $East_{raw}$ and $North_{raw}$ components are raw data after March 2011.

| | TOK1 | | | |
|---|---|---|---|---|
| Epoch | Eastward | Northward | $East_{raw}$ | $North_{raw}$ |
| (year) | (m) | (m) | (m) | (m) |
| 2006.63 | 0 | 0 | | |
| 2007.29 | 0.0313 | -0.0173 | | |
| 2008.54 | 0.0011 | -0.0051 | | |
| 2009.34 | -0.026 | 0.01 | | |
| 2009.55 | 0.0001 | -0.0035 | | |
| 2009.63 | -0.007 | -0.0094 | | |
| 2010.6 | -0.0818 | -0.0221 | | |
| 2010.95 | -0.0409 | -0.0016 | | |
| 2011.07 | -0.068 | -0.0099 | | |
| 2011.44 | -0.0594 | -0.0128 | 0.0024 | 0.0156 |
| 2011.94 | -0.0715 | -0.0285 | -0.0005 | 0.0133 |
| 2012.42 | -0.0517 | -0.0443 | 0.0232 | 0.0069 |
| 2012.71 | -0.0689 | -0.0368 | 0.0077 | 0.0194 |
| 2013.09 | -0.1327 | -0.0213 | -0.0539 | 0.0408 |
| 2013.4 | -0.1482 | -0.0461 | -0.0674 | 0.0205 |
| 2013.5 | -0.1252 | -0.0094 | -0.0437 | 0.0585 |
| 2013.88 | -0.1395 | -0.0256 | -0.0556 | 0.047 |
| 2014.69 | -0.1585 | -0.1137 | -0.0696 | -0.0335 |
| 2015.21 | -0.1733 | -0.0263 | -0.0813 | 0.0581 |

| | TOK2 | | | |
|---|---|---|---|---|
| Epoch | Eastward | Northward | $East_{raw}$ | $North_{raw}$ |
| (year) | (m) | (m) | (m) | (m) |
| 2006.85 | 0 | 0 | | |
| 2007.64 | 0.0008 | 0.0069 | | |
| 2008.03 | -0.0788 | -0.0163 | | |

| Epoch (year) | Eastward (m) | Northward (m) | East$_{raw}$ (m) | North$_{raw}$ (m) |
|---|---|---|---|---|
| 2008.53 | -0.0648 | -0.002 | | |
| 2008.53 | -0.0828 | -0.0132 | | |
| 2009.1 | -0.0999 | -0.0222 | | |
| 2009.55 | -0.109 | -0.0422 | | |
| 2010.06 | -0.1268 | 0.0106 | | |
| 2010.6 | -0.1071 | -0.0271 | | |
| 2010.96 | -0.1207 | -0.0099 | | |
| 2011.07 | -0.1272 | 0.001 | | |
| 2011.44 | -0.1839 | -0.005 | -0.1342 | 0.0233 |
| 2011.94 | -0.1852 | 0.0003 | -0.1279 | 0.0415 |
| 2012.77 | -0.2016 | -0.0223 | -0.139 | 0.0337 |
| 2013.41 | -0.1939 | -0.0236 | -0.1276 | 0.0414 |
| 2013.88 | -0.2347 | -0.0106 | -0.1656 | 0.0602 |
| 2014.07 | -0.2016 | -0.0227 | -0.1315 | 0.0502 |
| 2014.3 | -0.2033 | -0.0272 | -0.1319 | 0.0478 |
| 2015.2 | -0.2277 | -0.0401 | -0.1514 | 0.0425 |

| TOK3 | | | | |
|---|---|---|---|---|
| Epoch (year) | Eastward (m) | Northward (m) | East$_{raw}$ (m) | North$_{raw}$ (m) |
| 2012.1 | 0 | 0 | 0 | 0 |
| 2012.15 | -0.026 | -0.0179 | -0.0254 | -0.0166 |
| 2012.42 | -0.02 | 0.0206 | -0.0166 | 0.0276 |
| 2013.08 | 0.0019 | -0.0245 | 0.0109 | -0.0056 |
| 2013.44 | -0.0293 | -0.0047 | -0.0172 | 0.0199 |
| 2013.91 | -0.0494 | -0.0218 | -0.0332 | 0.0094 |
| 2014.29 | -0.0653 | 0.005 | -0.0459 | 0.0408 |
| 2015.17 | -0.0951 | -0.0165 | -0.0689 | 0.0279 |

| KUM1 | | | | |
|---|---|---|---|---|
| Epoch (year) | Eastward (m) | Northward (m) | East$_{raw}$ (m) | North$_{raw}$ (m) |
| 2007.64 | 0 | 0 | | |
| 2008.73 | 0.0246 | 0.0184 | | |
| 2009.34 | 0.0216 | 0.0203 | | |
| 2009.55 | 0.019 | 0.0011 | | |

| Epoch (year) | Eastward (m) | Northward (m) | East_raw_ (m) | North_raw_ (m) |
|---|---|---|---|---|
| 2009.83 | 0.0313 | -0.0108 | | |
| 2010.04 | 0.0623 | -0.0227 | | |
| 2010.6 | 0.0221 | 0.0253 | | |
| 2011.06 | 0.0113 | 0.0094 | | |
| 2011.44 | 0.022 | 0.0115 | 0.0607 | 0.0388 |
| 2012.77 | -0.0223 | 0.0021 | 0.027 | 0.0546 |
| 2013.41 | -0.0219 | -0.0204 | 0.0305 | 0.0404 |
| 2013.51 | -0.0139 | -0.0148 | 0.0391 | 0.0472 |
| 2014.07 | -0.047 | -0.0194 | 0.0088 | 0.0486 |
| 2014.69 | -0.0417 | -0.0037 | 0.017 | 0.0696 |
| 2015.2 | -0.0425 | -0.0642 | 0.0186 | 0.0131 |

| KUM2 | | | | |
|---|---|---|---|---|
| Epoch (year) | Eastward (m) | Northward (m) | East_raw_ (m) | North_raw_ (m) |
| 2012.09 | 0 | 0 | 0 | 0 |
| 2012.43 | -0.0027 | -0.0327 | -0.0012 | -0.0276 |
| 2013.07 | -0.0336 | 0.016 | -0.0295 | 0.0295 |
| 2013.43 | -0.0112 | -0.0186 | -0.0056 | -0.0011 |
| 2013.89 | -0.0238 | -0.0657 | -0.0163 | -0.0435 |
| 2014.3 | -0.0567 | -0.0174 | -0.0475 | 0.0084 |
| 2014.98 | -0.0458 | -0.0114 | -0.034 | 0.0194 |

| KUM3 | | | | |
|---|---|---|---|---|
| Epoch (year) | Eastward (m) | Northward (m) | East_raw_ (m) | North_raw_ (m) |
| 2006.41 | 0 | 0 | | |
| 2007.34 | 0.0535 | 0.021 | | |
| 2007.66 | 0.0097 | 0.0294 | | |
| 2008.71 | 0.0507 | 0.0434 | | |
| 2009.11 | 0.0416 | 0.0304 | | |
| 2009.33 | 0.0117 | 0.066 | | |
| 2009.55 | 0.0322 | 0.0303 | | |
| 2010.06 | 0.0089 | 0.0092 | | |
| 2010.6 | 0.0097 | 0.0137 | | |
| 2011.06 | 0.0159 | 0.018 | | |

| Epoch (year) | Eastward (m) | Northward (m) | East$_{raw}$ (m) | North$_{raw}$ (m) |
|---|---|---|---|---|
| 2011.44 | -0.0131 | 0.0133 | 0.014 | 0.0381 |
| 2011.9 | -0.0057 | 0.0277 | 0.0253 | 0.0619 |
| 2011.95 | -0.0255 | 0.0158 | 0.0057 | 0.0508 |
| 2012.8 | -0.0233 | -0.0203 | 0.0113 | 0.0264 |
| 2013.43 | -0.0502 | -0.03 | -0.0132 | 0.0238 |
| 2013.89 | -0.0513 | 0.0453 | -0.0125 | 0.1035 |
| 2014.69 | -0.0415 | 0.0006 | 0.0002 | 0.065 |
| 2015.18 | -0.0777 | 0.0414 | -0.0343 | 0.1093 |

| SIOW | | | | |
|---|---|---|---|---|
| Epoch (year) | Eastward (m) | Northward (m) | East$_{raw}$ (m) | North$_{raw}$ (m) |
| 2006.69 | 0 | 0 | | |
| 2007.65 | -0.0583 | -0.044 | | |
| 2008.04 | 0.0016 | -0.0475 | | |
| 2008.71 | -0.0133 | -0.0594 | | |
| 2009.32 | -0.0805 | -0.031 | | |
| 2009.55 | -0.05 | -0.04 | | |
| 2010.05 | -0.0604 | -0.0253 | | |
| 2010.6 | -0.0728 | -0.0092 | | |
| 2011.06 | -0.1209 | -0.0232 | | |
| 2011.44 | -0.0768 | -0.0094 | -0.055 | 0.0152 |
| 2011.95 | -0.0848 | -0.0125 | -0.0596 | 0.0214 |
| 2012.78 | -0.0811 | -0.0212 | -0.0529 | 0.0233 |
| 2013.06 | -0.1283 | -0.0246 | -0.0991 | 0.0229 |
| 2013.43 | -0.104 | -0.0208 | -0.0735 | 0.0303 |
| 2013.89 | -0.1684 | -0.0163 | -0.1363 | 0.0389 |
| 2014.68 | -0.1412 | 0.0097 | -0.1065 | 0.0707 |
| 2014.97 | -0.1927 | -0.0198 | -0.1571 | 0.0432 |

| MRT1 | | | | |
|---|---|---|---|---|
| Epoch (year) | Eastward (m) | Northward (m) | East$_{raw}$ (m) | North$_{raw}$ (m) |
| 2012.09 | 0 | 0 | 0 | 0 |
| 2012.16 | 0.0114 | 0.028 | 0.0117 | 0.0289 |
| 2012.44 | -0.0216 | -0.0093 | -0.0201 | -0.0048 |

| Epoch (year) | Eastward (m) | Northward (m) | East$_{raw}$ (m) | North$_{raw}$ (m) |
|---|---|---|---|---|
| 2012.78 | -0.0058 | 0.0049 | -0.0029 | 0.0133 |
| 2013.43 | -0.0253 | -0.0073 | -0.0196 | 0.0077 |
| 2013.9 | -0.0505 | -0.0241 | -0.0428 | -0.0049 |
| 2014.67 | -0.0221 | -0.008 | -0.0114 | 0.017 |
| 2014.97 | -0.0251 | 0.0033 | -0.0134 | 0.0304 |

| MRT2 | | | | |
|---|---|---|---|---|
| Epoch (year) | Eastward (m) | Northward (m) | East$_{raw}$ (m) | North$_{raw}$ (m) |
| 2006.67 | 0 | 0 | | |
| 2007.35 | 0.0084 | -0.043 | | |
| 2008.71 | 0.0184 | -0.0375 | | |
| 2009.11 | -0.0125 | -0.0438 | | |
| 2009.33 | 0.0071 | -0.0116 | | |
| 2009.56 | 0.0021 | -0.0562 | | |
| 2010.05 | 0.0252 | -0.0557 | | |
| 2010.6 | -0.0253 | -0.043 | | |
| 2010.96 | -0.0364 | -0.0619 | | |
| 2011.44 | -0.0056 | -0.0977 | 0.0098 | -0.0756 |
| 2011.95 | -0.0271 | -0.0627 | -0.0093 | -0.0326 |
| 2012.78 | -0.0568 | -0.0637 | -0.0368 | -0.0248 |
| 2013.06 | -0.0538 | -0.0593 | -0.0331 | -0.0179 |
| 2013.42 | -0.0256 | -0.0456 | -0.0039 | -0.0012 |
| 2013.89 | -0.0732 | -0.0693 | -0.0502 | -0.0213 |
| 2014.3 | -0.0767 | -0.0774 | -0.0527 | -0.0267 |
| 2014.67 | -0.0267 | -0.063 | -0.0019 | -0.0101 |
| 2014.95 | -0.0731 | -0.0122 | -0.0476 | 0.0423 |
| 2015.18 | -0.1112 | -0.032 | -0.0852 | 0.0238 |
| 2015.43 | -0.1069 | -0.0431 | -0.0804 | 0.0141 |

| TOS1 | | | | |
|---|---|---|---|---|
| Epoch (year) | Eastward (m) | Northward (m) | East$_{raw}$ (m) | North$_{raw}$ (m) |
| 2012.07 | 0 | 0 | 0 | 0 |
| 2012.44 | -0.03 | -0.0124 | -0.0291 | -0.0089 |
| 2012.79 | -0.0696 | 0.0233 | -0.0679 | 0.0296 |

| Epoch (year) | Eastward (m) | Northward (m) | East$_{raw}$ (m) | North$_{raw}$ (m) |
|---|---|---|---|---|
| 2013.07 | -0.0378 | -0.0204 | -0.0354 | -0.012 |
| 2013.52 | -0.0602 | 0.0129 | -0.0567 | 0.0245 |
| 2014.04 | -0.0174 | -0.0074 | -0.0127 | 0.0075 |
| 2014.67 | -0.054 | 0.0254 | -0.0481 | 0.0437 |
| 2014.95 | -0.0769 | 0.0353 | -0.0704 | 0.0551 |
| 2015.41 | -0.1072 | 0.046 | -0.0999 | 0.0682 |

| TOS2 | | | | |
|---|---|---|---|---|
| Epoch (year) | Eastward (m) | Northward (m) | East$_{raw}$ (m) | North$_{raw}$ (m) |
| 2011.9 | 0 | 0 | 0 | 0 |
| 2011.95 | -0.0188 | 0.0066 | -0.0187 | 0.0071 |
| 2013.06 | -0.0438 | 0.03 | -0.0421 | 0.0392 |
| 2013.42 | -0.0809 | 0.0087 | -0.0787 | 0.0202 |
| 2013.9 | -0.07 | 0.0527 | -0.067 | 0.067 |
| 2014.06 | -0.0576 | 0.0605 | -0.0544 | 0.0757 |
| 2014.31 | -0.0696 | 0.0443 | -0.066 | 0.0608 |
| 2014.95 | -0.0564 | 0.0083 | -0.0521 | 0.0277 |
| 2015.4 | -0.0648 | 0.0427 | -0.0599 | 0.0642 |

| ASZ1 | | | | |
|---|---|---|---|---|
| Epoch (year) | Eastward (m) | Northward (m) | East$_{raw}$ (m) | North$_{raw}$ (m) |
| 2012.08 | 0 | 0 | 0 | 0 |
| 2012.16 | -0.0133 | 0.0266 | -0.0132 | 0.0272 |
| 2012.46 | -0.0243 | 0.0081 | -0.0238 | 0.0109 |
| 2013.04 | -0.0383 | 0.0035 | -0.037 | 0.01 |
| 2013.51 | -0.0207 | -0.0052 | -0.0188 | -0.004 |
| 2014.04 | -0.0088 | 0.043 | -0.0061 | 0.0549 |
| 2014.32 | -0.0461 | 0.024 | -0.0431 | 0.0371 |
| 2014.95 | -0.0375 | 0.0087 | -0.0338 | 0.0245 |
| 2015.2 | -0.0652 | 0.0552 | -0.0613 | 0.072 |
| 2015.41 | -0.0769 | 0.0246 | -0.0728 | 0.0423 |

| ASZ2 |
|---|

| Epoch (year) | Eastward (m) | Northward (m) | East$_{raw}$ (m) | North$_{raw}$ (m) |
|---|---|---|---|---|
| 2012.08 | 0 | 0 | 0 | 0 |
| 2012.09 | 0.0102 | 0.0038 | 0.0102 | 0.0039 |
| 2012.46 | 0.0124 | -0.0117 | 0.0126 | -0.0094 |
| 2013.03 | 0.0502 | -0.002 | 0.0507 | 0.0033 |
| 2013.51 | 0.0422 | -0.0009 | 0.0431 | 0.0066 |
| 2014.04 | -0.0121 | 0.0178 | -0.0108 | 0.0275 |
| 2014.67 | 0.0005 | -0.0247 | 0.0021 | -0.0127 |
| 2015.2 | -0.0444 | 0.0276 | -0.0426 | 0.0414 |
| 2015.4 | -0.0343 | 0.0113 | -0.0324 | 0.0258 |

| HYG1 | | | | |
|---|---|---|---|---|
| Epoch (year) | Eastward (m) | Northward (m) | East$_{raw}$ (m) | North$_{raw}$ (m) |
| 2012.08 | 0 | 0 | 0 | 0 |
| 2012.16 | -0.0066 | 0.0177 | -0.0065 | 0.0183 |
| 2012.45 | -0.0032 | 0.0123 | -0.0027 | 0.0148 |
| 2013.04 | -0.0304 | 0.0245 | -0.0292 | 0.0305 |
| 2013.52 | -0.0061 | 0.0165 | -0.0043 | 0.0249 |
| 2014.05 | -0.0427 | 0.0278 | -0.0403 | 0.0387 |
| 2014.32 | -0.0061 | 0.0575 | -0.0034 | 0.0695 |
| 2014.66 | 0.0191 | 0.0368 | 0.0221 | 0.0502 |
| 2014.94 | -0.0168 | 0.0251 | -0.0136 | 0.0396 |
| 2015.41 | -0.0339 | 0.0383 | -0.0303 | 0.0546 |

| HYG2 | | | | |
|---|---|---|---|---|
| Epoch (year) | Eastward (m) | Northward (m) | East$_{raw}$ (m) | North$_{raw}$ (m) |
| 2012.07 | 0 | 0 | 0 | 0 |
| 2012.17 | 0.0622 | -0.0368 | 0.0622 | -0.0362 |
| 2012.45 | -0.0006 | 0.0333 | -0.0004 | 0.0354 |
| 2013.05 | 0.0215 | 0.0008 | 0.0221 | 0.0057 |
| 2013.52 | 0.0333 | -0.0266 | 0.0343 | -0.0197 |
| 2014.05 | 0.0314 | -0.0366 | 0.0328 | -0.0277 |
| 2014.32 | 0.0258 | 0.0168 | 0.0274 | 0.0266 |
| 2014.66 | 0.0118 | 0.0035 | 0.0135 | 0.0144 |
| 2014.94 | 0.0665 | -0.0446 | 0.0684 | -0.0328 |
| 2015.41 | 0.0127 | -0.0308 | 0.0148 | -0.0176 |

**Supplementary Table 2 | Position and calculated SDR value of each subfault in the inversion analysis.** First and second columns show location (longitude and latitude) of each subfault. Third and fourth columns show angle (from the east to the counterclockwise direction) and absolute SDR value (m/year) calculated for each subfault in the inversion.

| LON(E) | LAT(N) | ANGLE | ABSOLUTE VALUE[m/year] |
| --- | --- | --- | --- |
| 132.511 | 31.041 | −143.373 | 0.0077 |
| 132.743 | 31.169 | −133.469 | 0.0077 |
| 132.976 | 31.296 | −135.328 | 0.0083 |
| 133.21 | 31.422 | −150.873 | 0.0121 |
| 133.445 | 31.549 | −161.94 | 0.0189 |
| 133.68 | 31.675 | −168.515 | 0.0233 |
| 133.916 | 31.8 | −175.832 | 0.022 |
| 134.153 | 31.925 | 170.747 | 0.0162 |
| 134.39 | 32.049 | 153.338 | 0.0117 |
| 134.628 | 32.173 | 145.007 | 0.0097 |
| 134.866 | 32.297 | 150.987 | 0.0089 |
| 135.106 | 32.42 | 165.316 | 0.0087 |
| 135.346 | 32.542 | 171.046 | 0.0099 |
| 135.586 | 32.664 | 162.653 | 0.0131 |
| 135.828 | 32.785 | 160.316 | 0.0153 |
| 136.07 | 32.906 | 160.591 | 0.0153 |
| 136.313 | 33.027 | 157.388 | 0.0154 |
| 136.557 | 33.146 | 160.04 | 0.0182 |
| 136.801 | 33.266 | 164.781 | 0.0191 |
| 137.046 | 33.384 | 170.011 | 0.0143 |
| 137.292 | 33.503 | 179.921 | 0.0178 |
| 137.539 | 33.62 | −176.997 | 0.0265 |
| 137.786 | 33.737 | −177.636 | 0.0278 |
| 138.035 | 33.854 | 176.539 | 0.0256 |
| 138.284 | 33.97 | 169.738 | 0.0225 |
| 138.533 | 34.085 | 168.921 | 0.017 |
| 138.784 | 34.2 | 164.725 | 0.0084 |
| 139.035 | 34.314 | 90.647 | 0.0014 |
| 139.288 | 34.427 | −3.338 | 0.0033 |
| 132.362 | 31.24 | −145.816 | 0.0255 |
| 132.595 | 31.368 | −135.005 | 0.0245 |
| 132.829 | 31.496 | −135.665 | 0.025 |
| 133.063 | 31.623 | −151.855 | 0.036 |
| 133.298 | 31.749 | −163.251 | 0.0567 |
| 133.533 | 31.876 | −170.083 | 0.0705 |
| 133.769 | 32.001 | −177.902 | 0.0674 |
| 134.006 | 32.126 | 168.09 | 0.051 |
| 134.244 | 32.251 | 151.463 | 0.0383 |
| 134.482 | 32.375 | 144.659 | 0.0324 |
| 134.721 | 32.499 | 150.978 | 0.0298 |
| 134.961 | 32.622 | 165.319 | 0.0281 |
| 135.201 | 32.745 | 171.167 | 0.0311 |
| 135.443 | 32.866 | 161.839 | 0.0408 |
| 135.685 | 32.988 | 159.047 | 0.0474 |
| 135.927 | 33.108 | 159.828 | 0.047 |
| 136.171 | 33.229 | 157.805 | 0.0461 |
| 136.415 | 33.349 | 160.505 | 0.0543 |
| 136.66 | 33.468 | 164.631 | 0.0571 |
| 136.905 | 33.588 | 169.406 | 0.042 |
| 137.151 | 33.707 | 179.176 | 0.0518 |
| 137.398 | 33.825 | −177.917 | 0.0774 |
| 137.646 | 33.942 | −178.735 | 0.0816 |
| 137.894 | 34.059 | 175.611 | 0.0757 |
| 138.144 | 34.175 | 169.234 | 0.067 |
| 138.394 | 34.291 | 168.515 | 0.0508 |
| 138.645 | 34.406 | 164.35 | 0.0255 |
| 138.897 | 34.52 | 96.337 | 0.004 |
| 139.149 | 34.634 | −8.721 | 0.0094 |
| 132.214 | 31.439 | −153.706 | 0.04 |
| 132.447 | 31.567 | −141.557 | 0.0327 |
| 132.681 | 31.695 | −138.542 | 0.0266 |
| 132.915 | 31.822 | −158.056 | 0.036 |
| 133.15 | 31.949 | −170.29 | 0.0598 |
| 133.386 | 32.076 | −177.761 | 0.078 |
| 133.622 | 32.202 | 172.861 | 0.0794 |
| 133.86 | 32.327 | 158.047 | 0.0679 |
| 134.098 | 32.452 | 145.644 | 0.0581 |
| 134.336 | 32.576 | 143.731 | 0.0517 |
| 134.576 | 32.7 | 150.75 | 0.0472 |
| 134.816 | 32.823 | 164.68 | 0.041 |
| 135.057 | 32.946 | 170.857 | 0.0404 |
| 135.299 | 33.068 | 158.148 | 0.0512 |
| 135.542 | 33.189 | 153.699 | 0.0591 |
| 135.785 | 33.309 | 156.285 | 0.0557 |
| 136.029 | 33.429 | 159.336 | 0.0492 |
| 136.274 | 33.549 | 162.093 | 0.0571 |
| 136.519 | 33.669 | 163.043 | 0.0607 |
| 136.764 | 33.79 | 165.244 | 0.0412 |
| 137.01 | 33.91 | 173.97 | 0.0471 |
| 137.257 | 34.029 | 176.406 | 0.0723 |
| 137.505 | 34.146 | 174.91 | 0.0782 |
| 137.754 | 34.263 | 170.37 | 0.0753 |
| 138.004 | 34.379 | 166.309 | 0.0687 |
| 138.255 | 34.496 | 166.463 | 0.0536 |
| 138.506 | 34.611 | 162.994 | 0.0286 |
| 138.758 | 34.726 | 135.172 | 0.0047 |
| 139.01 | 34.841 | −42.067 | 0.0083 |
| 132.065 | 31.637 | −166.816 | 0.0475 |
| 132.298 | 31.766 | −163.192 | 0.0323 |
| 132.532 | 31.894 | −174.867 | 0.0177 |
| 132.767 | 32.021 | 162.834 | 0.0238 |
| 133.002 | 32.149 | 162.798 | 0.0403 |
| 133.238 | 32.275 | 160.74 | 0.0568 |
| 133.475 | 32.401 | 154.144 | 0.0671 |
| 133.713 | 32.527 | 145.565 | 0.0698 |
| 133.952 | 32.652 | 142.409 | 0.068 |
| 134.191 | 32.776 | 145.242 | 0.064 |
| 134.43 | 32.9 | 150.023 | 0.0586 |
| 134.671 | 33.023 | 158.777 | 0.047 |
| 134.913 | 33.146 | 162.578 | 0.0396 |
| 135.156 | 33.267 | 148.687 | 0.047 |
| 135.4 | 33.387 | 143.108 | 0.0526 |
| 135.644 | 33.507 | 148.163 | 0.0448 |
| 135.889 | 33.627 | 163.769 | 0.0336 |
| 136.134 | 33.748 | 167.969 | 0.0395 |
| 136.379 | 33.869 | 161.226 | 0.0449 |
| 136.624 | 33.99 | 156.246 | 0.029 |
| 136.869 | 34.112 | 154.455 | 0.0247 |
| 137.116 | 34.232 | 155.292 | 0.0343 |
| 137.365 | 34.349 | 153.674 | 0.0389 |
| 137.615 | 34.465 | 155.544 | 0.0412 |
| 137.866 | 34.581 | 160.16 | 0.042 |
| 138.117 | 34.697 | 163.817 | 0.0386 |
| 138.368 | 34.814 | 162.392 | 0.0294 |
| 138.619 | 34.93 | 165.36 | 0.0186 |
| 138.871 | 35.047 | −160.699 | 0.0152 |
| 131.916 | 31.835 | 179.363 | 0.0513 |
| 132.148 | 31.965 | 167.627 | 0.0404 |
| 132.383 | 32.093 | 142.174 | 0.0344 |
| 132.618 | 32.221 | 132.343 | 0.041 |
| 132.854 | 32.348 | 137.297 | 0.0452 |
| 133.091 | 32.474 | 142.562 | 0.0539 |
| 133.328 | 32.6 | 143.852 | 0.0654 |
| 133.566 | 32.726 | 144.357 | 0.0729 |
| 133.805 | 32.851 | 147.184 | 0.0753 |
| 134.045 | 32.975 | 149.871 | 0.0739 |
| 134.285 | 33.1 | 148.929 | 0.0686 |
| 134.526 | 33.223 | 148.905 | 0.0559 |
| 134.769 | 33.344 | 147.982 | 0.0456 |
| 135.014 | 33.464 | 141.284 | 0.0482 |
| 135.259 | 33.583 | 137.763 | 0.0485 |
| 135.504 | 33.703 | 144.268 | 0.0387 |
| 135.75 | 33.823 | 170.472 | 0.0309 |
| 135.995 | 33.944 | 175.116 | 0.0412 |
| 136.24 | 34.066 | 165.232 | 0.0507 |
| 136.484 | 34.188 | 160.814 | 0.041 |
| 136.729 | 34.312 | 156.037 | 0.0311 |
| 136.975 | 34.433 | 149.24 | 0.0284 |
| 137.225 | 34.55 | 145.077 | 0.028 |
| 137.477 | 34.664 | 152.655 | 0.0305 |
| 137.729 | 34.779 | 163.814 | 0.0369 |
| 137.981 | 34.895 | 168.361 | 0.0438 |
| 138.232 | 35.012 | 164.854 | 0.0475 |
| 138.484 | 35.129 | 162.993 | 0.0483 |
| 138.735 | 35.248 | 172.213 | 0.0476 |
| 131.769 | 32.028 | 173.113 | 0.0446 |
| 131.999 | 32.162 | 156.953 | 0.0492 |
| 132.233 | 32.292 | 141.669 | 0.0558 |
| 132.468 | 32.42 | 137.858 | 0.0604 |
| 132.704 | 32.547 | 142.42 | 0.0532 |
| 132.942 | 32.674 | 146.413 | 0.0522 |
| 133.18 | 32.799 | 146.988 | 0.0606 |
| 133.419 | 32.924 | 148.962 | 0.0699 |
| 133.659 | 33.049 | 151.73 | 0.0753 |
| 133.899 | 33.174 | 152.055 | 0.0752 |
| 134.139 | 33.298 | 147.88 | 0.0698 |
| 134.381 | 33.422 | 145.399 | 0.0579 |
| 134.625 | 33.542 | 147.418 | 0.0458 |
| 134.872 | 33.659 | 150.995 | 0.0407 |
| 135.12 | 33.776 | 152.511 | 0.0341 |
| 135.367 | 33.894 | 161.331 | 0.0244 |
| 135.613 | 34.014 | −169.428 | 0.0221 |
| 135.858 | 34.135 | −174.42 | 0.0322 |
| 136.103 | 34.258 | 170.717 | 0.0436 |
| 136.347 | 34.383 | 166.917 | 0.0435 |
| 136.589 | 34.51 | 164.585 | 0.038 |
| 136.834 | 34.634 | 158.084 | 0.0317 |
| 137.085 | 34.75 | 151.575 | 0.0257 |
| 137.339 | 34.863 | 158.435 | 0.0248 |
| 137.592 | 34.976 | 170.922 | 0.0318 |
| 137.847 | 35.089 | 173.824 | 0.0437 |
| 138.102 | 35.202 | 166.972 | 0.0555 |
| 138.355 | 35.318 | 161.692 | 0.0646 |
| 138.606 | 35.438 | 165.788 | 0.0675 |
| 131.63 | 32.209 | −175.178 | 0.0288 |
| 131.853 | 32.354 | 160.225 | 0.0468 |
| 132.083 | 32.489 | 147.516 | 0.0674 |
| 132.318 | 32.619 | 145.41 | 0.0732 |
| 132.554 | 32.747 | 151.259 | 0.0572 |
| 132.793 | 32.873 | 154.85 | 0.0463 |
| 133.032 | 32.998 | 151.486 | 0.0492 |
| 133.272 | 33.122 | 151.081 | 0.0584 |
| 133.512 | 33.247 | 151.653 | 0.0643 |
| 133.752 | 33.372 | 150.062 | 0.0641 |
| 133.993 | 33.497 | 145.897 | 0.0594 |
| 134.236 | 33.62 | 146.064 | 0.0484 |
| 134.483 | 33.737 | 156.882 | 0.035 |
| 134.733 | 33.851 | 178.464 | 0.027 |
| 134.983 | 33.964 | −163.278 | 0.0205 |
| 135.231 | 34.081 | −141.558 | 0.0136 |
| 135.477 | 34.202 | −115.701 | 0.0124 |
| 135.722 | 34.325 | −141.512 | 0.0132 |
| 135.967 | 34.449 | −174.858 | 0.0215 |
| 136.21 | 34.575 | −179.236 | 0.0304 |
| 136.45 | 34.705 | 177.877 | 0.035 |
| 136.693 | 34.834 | 169.185 | 0.0331 |
| 136.946 | 34.948 | 161.35 | 0.0248 |
| 137.202 | 35.059 | 166.28 | 0.0183 |
| 137.457 | 35.17 | −178.008 | 0.0202 |
| 137.716 | 35.279 | −176.851 | 0.0305 |
| 137.975 | 35.386 | 170.653 | 0.0444 |
| 138.232 | 35.497 | 161.273 | 0.0574 |
| 138.484 | 35.615 | 162.266 | 0.0621 |
| 131.505 | 32.373 | −125.698 | 0.0288 |
| 131.717 | 32.532 | −177.212 | 0.039 |
| 131.939 | 32.679 | 154.713 | 0.066 |
| 132.169 | 32.815 | 146.718 | 0.0766 |
| 132.404 | 32.945 | 147.015 | 0.0571 |
| 132.643 | 33.071 | 146.757 | 0.0389 |
| 132.883 | 33.196 | 144.008 | 0.0365 |
| 133.124 | 33.32 | 146.72 | 0.0427 |
| 133.365 | 33.444 | 148.441 | 0.0454 |
| 133.606 | 33.569 | 146.811 | 0.0438 |
| 133.846 | 33.695 | 144.58 | 0.0417 |
| 134.09 | 33.818 | 148.035 | 0.0328 |
| 134.34 | 33.931 | 168.965 | 0.0194 |
| 134.596 | 34.038 | −139.646 | 0.0138 |
| 134.85 | 34.147 | −110.622 | 0.0122 |
| 135.1 | 34.263 | −90.404 | 0.007 |
| 135.345 | 34.385 | −42.584 | 0.0051 |
| 135.589 | 34.51 | −59.877 | 0.0042 |
| 135.833 | 34.635 | −130.869 | 0.0103 |
| 136.075 | 34.764 | −145.393 | 0.0241 |
| 136.312 | 34.9 | −156.112 | 0.0355 |
| 136.551 | 35.034 | −171.422 | 0.039 |
| 136.806 | 35.146 | 178.998 | 0.0308 |
| 137.067 | 35.251 | −176.881 | 0.0186 |
| 137.326 | 35.358 | −152.055 | 0.0119 |
| 137.588 | 35.462 | −144.708 | 0.0146 |
| 137.852 | 35.563 | −171.142 | 0.0196 |
| 138.115 | 35.667 | 166.033 | 0.027 |
| 138.373 | 35.778 | 161.08 | 0.031 |
| 131.394 | 32.517 | −108.03 | 0.0599 |
| 131.595 | 32.691 | −139.019 | 0.0546 |
| 131.805 | 32.854 | 178.188 | 0.063 |
| 132.025 | 33.004 | 158.884 | 0.074 |
| 132.255 | 33.141 | 154.015 | 0.0601 |
| 132.493 | 33.269 | 153.311 | 0.0429 |
| 132.734 | 33.393 | 156.196 | 0.0364 |
| 132.977 | 33.515 | 164.363 | 0.0368 |
| 133.219 | 33.639 | 171.855 | 0.034 |
| 133.46 | 33.765 | 175.483 | 0.0285 |
| 133.699 | 33.892 | 173.862 | 0.0261 |
| 133.942 | 34.016 | 177.879 | 0.0202 |
| 134.199 | 34.122 | −152.679 | 0.0109 |
| 134.463 | 34.218 | −81.386 | 0.0095 |
| 134.723 | 34.32 | −40.108 | 0.0062 |
| 134.976 | 34.433 | 74.93 | 0.0065 |
| 135.223 | 34.554 | 91.232 | 0.0125 |
| 135.466 | 34.68 | 113.644 | 0.0091 |
| 135.709 | 34.807 | −157.56 | 0.0136 |
| 135.947 | 34.942 | −141.394 | 0.0335 |
| 136.176 | 35.09 | −144.929 | 0.0499 |
| 136.408 | 35.235 | −161.078 | 0.0558 |
| 136.668 | 35.341 | −171.413 | 0.0435 |
| 136.935 | 35.437 | −162.489 | 0.0213 |
| 137.201 | 35.536 | −75.845 | 0.01 |
| 137.47 | 35.631 | −42.007 | 0.0195 |
| 137.742 | 35.721 | −31.726 | 0.0193 |
| 138.014 | 35.812 | −9.051 | 0.014 |
| 138.281 | 35.911 | 24.465 | 0.0116 |

**Supplementary Table 3 | Observed and calculated vectors of each site in the inversion analysis.** First and second columns show location (longitude and latitude) of each site. Third and fourth columns show angle (from the east to the counterclockwise direction) and absolute velocity value (m/year) observed in each site. Fifth and sixth columns show angle and absolute velocity value calculated for each site in the inversion.

| LON(E) | LAT(N) | obs(ANG, | ABS[m/y]) | calc(ANG, | ABS[m/y]) |
|---:|---:|---:|---:|---:|---:|
| 135.4 | 33.745 | 164.646 | 0.0279 | 163.81 | 0.028 |
| 136.338 | 34.65 | 174.75 | 0.0197 | 176.701 | 0.0186 |
| 131.798 | 33.24 | 166.176 | 0.0209 | 164.742 | 0.0205 |
| 133.682 | 33.938 | 149.24 | 0.0217 | 150.588 | 0.0215 |
| 132.458 | 34.894 | 144.851 | 0.0048 | 138.315 | 0.0052 |
| 136.165 | 35.096 | 175.447 | 0.0156 | 179.733 | 0.0159 |
| 134.395 | 34.219 | 151.181 | 0.0144 | 149.485 | 0.0144 |
| 132.195 | 34.371 | 140.015 | 0.0086 | 141.4 | 0.0073 |
| 135.77 | 33.484 | 165.501 | 0.0369 | 164.757 | 0.0368 |
| 135.761 | 35.488 | 175.315 | 0.0091 | 161.152 | 0.0082 |
| 138.118 | 34.68 | 166.015 | 0.03 | 165.542 | 0.0306 |
| 131.365 | 32.528 | 175.798 | 0.0186 | 179.689 | 0.0151 |
| 135.389 | 33.962 | 166.049 | 0.0246 | 166.933 | 0.0227 |
| 137.165 | 34.918 | 166.522 | 0.0244 | 167.154 | 0.0239 |
| 132.991 | 34.079 | 144.582 | 0.0144 | 140.846 | 0.0142 |
| 135.714 | 33.638 | 167.329 | 0.0311 | 166.102 | 0.0311 |
| 131.149 | 32.457 | −168.614 | 0.0119 | −173.386 | 0.0134 |
| 136.862 | 34.466 | 163.368 | 0.0245 | 163.604 | 0.0264 |
| 133.806 | 33.655 | 151.003 | 0.0298 | 151.854 | 0.0305 |
| 131.416 | 34.441 | 137.085 | 0.0051 | 146.598 | 0.0049 |
| 134.237 | 35.022 | 155.266 | 0.0059 | 150.109 | 0.0062 |
| 131.066 | 34.181 | 143.898 | 0.0051 | 150.477 | 0.0049 |
| 135.74 | 33.97 | 169.276 | 0.0234 | 172.397 | 0.023 |
| 137.352 | 34.787 | 165.712 | 0.0254 | 163.822 | 0.0255 |
| 133.193 | 35.313 | 144.613 | 0.0031 | 137.542 | 0.0044 |
| 138.179 | 34.634 | 164.994 | 0.0322 | 167.227 | 0.0327 |
| 131.384 | 33.079 | 177.419 | 0.0171 | 170.824 | 0.015 |
| 132.903 | 35.312 | 94.142 | 0.0029 | 137.115 | 0.0043 |
| 138.036 | 34.786 | 166.515 | 0.0275 | 165.135 | 0.0275 |
| 137.327 | 35.04 | 166.407 | 0.0226 | 168.292 | 0.023 |
| 135.877 | 35.138 | 173.902 | 0.0128 | 174.688 | 0.0127 |
| 131.531 | 32.557 | 174.548 | 0.0163 | 173.257 | 0.0163 |
| 133.901 | 34.327 | 143.218 | 0.0117 | 144.79 | 0.0119 |
| 135.595 | 33.748 | 166.927 | 0.0279 | 166.734 | 0.0276 |
| 135.671 | 34.99 | 169.514 | 0.0124 | 170.96 | 0.0127 |
| 137.291 | 34.649 | 165.953 | 0.0262 | 160.715 | 0.0269 |
| 133.962 | 34.984 | 154.705 | 0.0059 | 145.379 | 0.0059 |
| 134.662 | 34.866 | 157.786 | 0.0075 | 158.968 | 0.0079 |
| 136.549 | 34.286 | 174.342 | 0.0237 | 168.754 | 0.0251 |
| 132.487 | 33.612 | 153.855 | 0.0214 | 152.178 | 0.0209 |
| 132.885 | 34.211 | 140.213 | 0.0119 | 138.489 | 0.0114 |
| 131.132 | 32.845 | −173.677 | 0.0148 | 173.544 | 0.0129 |
| 135.632 | 34.48 | 169.355 | 0.0178 | 171.891 | 0.0169 |
| 138.271 | 34.954 | 168.528 | 0.0263 | 168.393 | 0.0262 |
| 133.34 | 35.438 | 120.361 | 0.0028 | 137.931 | 0.0042 |
| 133.876 | 34.041 | 142.28 | 0.0191 | 148.874 | 0.0192 |
| 135.184 | 33.915 | 159.933 | 0.0244 | 164.447 | 0.0234 |
| 131.093 | 32.951 | −165.676 | 0.014 | 171.308 | 0.0119 |
| 132.281 | 33.47 | 158.443 | 0.0218 | 156.772 | 0.0215 |
| 133.338 | 34.779 | 144.965 | 0.0065 | 137.579 | 0.0065 |
| 133.802 | 34.739 | 146.976 | 0.0064 | 142.471 | 0.0068 |
| 133 | 32.992 | 153.603 | 0.0396 | 153.261 | 0.0399 |
| 134.738 | 34.335 | 154.364 | 0.0132 | 154.884 | 0.0128 |
| 133.281 | 33.469 | 153.651 | 0.0332 | 152.392 | 0.0322 |
| 134.158 | 35.342 | 159.677 | 0.0049 | 145.167 | 0.0052 |
| 134.774 | 35.164 | 165.357 | 0.0066 | 154.198 | 0.0073 |
| 137.711 | 35.08 | 167.954 | 0.0229 | 170.041 | 0.0223 |
| 135.225 | 34.159 | 165.156 | 0.0198 | 163.53 | 0.0185 |
| 131.169 | 33.497 | 159.739 | 0.0093 | 158.743 | 0.0087 |
| 135.362 | 34.257 | 162.49 | 0.0186 | 166.136 | 0.0178 |
| 135.92 | 35.353 | 173.772 | 0.0107 | 171.64 | 0.0106 |
| 132.489 | 33.04 | 157.924 | 0.0317 | 153.1 | 0.0345 |
| 132.361 | 33.203 | 154.487 | 0.028 | 159.162 | 0.0293 |
| 137.086 | 35.24 | 169.114 | 0.0212 | 171.004 | 0.0203 |
| 133.565 | 35.171 | 147.06 | 0.004 | 139.784 | 0.005 |
| 134.239 | 35.268 | 159.529 | 0.0055 | 147.05 | 0.0055 |
| 135.691 | 34.779 | 174.374 | 0.0161 | 172.687 | 0.0148 |
| 134.628 | 33.928 | 157.505 | 0.0236 | 160.218 | 0.0219 |
| 134.976 | 35.092 | 160.802 | 0.0085 | 157.861 | 0.0083 |
| 134.101 | 33.498 | 154.277 | 0.0386 | 154.003 | 0.0378 |
| 132.744 | 33.306 | 156.395 | 0.0293 | 157.801 | 0.0306 |
| 134.051 | 35.177 | 159.96 | 0.0052 | 145.676 | 0.0054 |
| 138.206 | 35.322 | 173.441 | 0.0214 | 171.997 | 0.0233 |
| 138.275 | 35.201 | 168.883 | 0.0268 | 171.071 | 0.0271 |
| 137.944 | 35.318 | 167.393 | 0.0214 | 171.72 | 0.0199 |
| 133.09 | 33.417 | 153.419 | 0.0316 | 154.115 | 0.0318 |
| 134.821 | 35.317 | 161.679 | 0.0064 | 150.805 | 0.0068 |
| 133.048 | 33.885 | 149.047 | 0.0191 | 146.986 | 0.0192 |
| 135.069 | 34.273 | 153.992 | 0.0147 | 160.538 | 0.0156 |
| 132.358 | 33.911 | 145.689 | 0.0141 | 143.474 | 0.0131 |
| 133.599 | 34.811 | 140.026 | 0.0065 | 139.812 | 0.0064 |
| 132.576 | 34.342 | 135.326 | 0.0099 | 138.726 | 0.0088 |
| 138.157 | 34.758 | 166.346 | 0.028 | 165.778 | 0.0277 |
| 134.327 | 35.363 | 153.133 | 0.0051 | 146.469 | 0.0054 |
| 134.487 | 33.823 | 157.261 | 0.0263 | 160.343 | 0.0251 |
| 133.794 | 35.265 | 158.392 | 0.0044 | 141.841 | 0.0049 |
| 137.809 | 34.793 | 166.307 | 0.0275 | 164.45 | 0.0265 |
| 132.053 | 34.091 | 143.229 | 0.0104 | 143.978 | 0.0088 |
| 133.927 | 34.165 | 146.536 | 0.0155 | 146.966 | 0.0158 |
| 136.152 | 34.988 | 173.892 | 0.0166 | 178.351 | 0.0164 |
| 136.842 | 34.905 | 170.072 | 0.022 | 170.165 | 0.0221 |
| 133.657 | 33.768 | 150.79 | 0.0255 | 151.781 | 0.026 |
| 131.47 | 32.022 | −179.716 | 0.0101 | −167.838 | 0.01 |
| 132.938 | 33.57 | 153.514 | 0.0259 | 156.903 | 0.0266 |
| 135.168 | 35.294 | 161.011 | 0.0082 | 154.037 | 0.0079 |
| 132.704 | 32.841 | 153.5 | 0.0391 | 153.647 | 0.0386 |
| 132.014 | 34.572 | 136.172 | 0.0055 | 141.928 | 0.0057 |
| 131.61 | 34.618 | 133.514 | 0.0049 | 144.462 | 0.0047 |
| 135.025 | 35.24 | 163.28 | 0.0078 | 154.203 | 0.0078 |
| 135.84 | 33.521 | 168.693 | 0.0341 | 166.602 | 0.0357 |
| 131.334 | 32.385 | −175.423 | 0.013 | −173.937 | 0.014 |
| 132.909 | 34.635 | 138.414 | 0.007 | 136.684 | 0.0071 |
| 133.129 | 33.216 | 152.27 | 0.0372 | 152.252 | 0.0376 |
| 136.799 | 35.302 | 173.503 | 0.0186 | 174.184 | 0.0183 |
| 134.316 | 34.474 | 148.015 | 0.0089 | 150.203 | 0.0097 |
| 137.591 | 35.101 | 166.349 | 0.0228 | 169.512 | 0.0221 |
| 131.254 | 32.002 | −156.817 | 0.0094 | −156.479 | 0.0078 |
| 135.33 | 34.978 | 159.634 | 0.0099 | 165.401 | 0.0107 |
| 131.531 | 34.058 | 147.378 | 0.0079 | 148.157 | 0.0067 |
| 135.486 | 34.528 | 167.708 | 0.016 | 170.212 | 0.0156 |
| 132.922 | 35.494 | 121.091 | 0.0023 | 136.959 | 0.0039 |
| 136.873 | 34.985 | 170.858 | 0.022 | 170.666 | 0.0218 |
| 131.176 | 34.343 | 156.828 | 0.0057 | 148.669 | 0.0047 |
| 133.404 | 33.408 | 153.147 | 0.0356 | 151.278 | 0.0356 |
| 133.648 | 34.069 | 148.5 | 0.0178 | 148.463 | 0.0177 |
| 131.564 | 33.672 | 150.433 | 0.0116 | 151.888 | 0.0101 |
| 134 | 34.447 | 140.58 | 0.0089 | 145.24 | 0.0097 |
| 136.205 | 34.061 | 168.677 | 0.0251 | 166.902 | 0.0255 |
| 135.385 | 34.726 | 173.034 | 0.0139 | 169.796 | 0.013 |
| 132.996 | 35.197 | 137.657 | 0.005 | 137.141 | 0.0046 |
| 131.516 | 32.171 | 178.15 | 0.013 | −178.34 | 0.0128 |
| 135.624 | 34.622 | 171.381 | 0.0161 | 172.184 | 0.0157 |
| 133.44 | 35.347 | 157.362 | 0.0048 | 138.618 | 0.0045 |
| 133.798 | 34.44 | 142.469 | 0.0102 | 142.87 | 0.0097 |
| 132.12 | 33.847 | 145.536 | 0.0138 | 145.265 | 0.012 |
| 134.378 | 34.762 | 156.446 | 0.007 | 154.527 | 0.0074 |
| 137.999 | 35.032 | 169.932 | 0.0238 | 170.142 | 0.0245 |
| 133.363 | 33.705 | 151.318 | 0.0251 | 153.028 | 0.0259 |
| 133.342 | 34.614 | 139.772 | 0.0082 | 137.624 | 0.0076 |
| 131.121 | 33.271 | 168.295 | 0.0097 | 164.277 | 0.01 |
| 136.638 | 34.432 | 169.252 | 0.0238 | 169.23 | 0.0244 |
| 138.642 | 35.203 | 177.439 | 0.0284 | 174.167 | 0.0292 |
| 138.094 | 34.94 | 167.427 | 0.0255 | 168.129 | 0.0259 |
| 132.694 | 33.447 | 166.717 | 0.0295 | 160.783 | 0.0271 |
| 132.974 | 33.313 | 153.544 | 0.033 | 154.561 | 0.033 |
| 132.475 | 33.385 | 156.25 | 0.025 | 162.553 | 0.0257 |
| 134.547 | 34.899 | 163.874 | 0.007 | 156.825 | 0.0074 |
| 136.351 | 34.209 | 169.201 | 0.0242 | 169.693 | 0.0241 |
| 132.54 | 34.68 | 150.442 | 0.0069 | 138.088 | 0.0063 |
| 134.864 | 34.494 | 156.06 | 0.0104 | 159.785 | 0.0112 |
| 135.559 | 33.942 | 169.226 | 0.0235 | 169.796 | 0.0232 |
| 133.19 | 34.253 | 141.66 | 0.0115 | 139.61 | 0.0116 |
| 132.821 | 34.336 | 142.615 | 0.008 | 137.629 | 0.0096 |
| 136.396 | 34.299 | 170.407 | 0.0237 | 171.774 | 0.0231 |
| 134.303 | 33.79 | 155.478 | 0.0275 | 157.123 | 0.0265 |
| 135.506 | 34.839 | 171.992 | 0.0126 | 170.278 | 0.013 |
| 131.757 | 32.704 | 163.553 | 0.0209 | 163.654 | 0.0226 |
| 135.877 | 33.8 | 172.601 | 0.0262 | 169.894 | 0.0275 |
| 135.176 | 34.688 | 166.04 | 0.0127 | 167.359 | 0.0116 |
| 134.948 | 34.938 | 157.97 | 0.0088 | 161.502 | 0.0089 |
| 133.701 | 35.492 | 135.554 | 0.0029 | 139.774 | 0.0043 |
| 138.065 | 34.666 | 165.212 | 0.0299 | 165.136 | 0.0313 |
| 137.053 | 35.129 | 167.977 | 0.0213 | 170.439 | 0.0215 |
| 132.903 | 35.037 | 152.024 | 0.0045 | 137 | 0.0051 |
| 132.104 | 33.976 | 146.798 | 0.0124 | 144.346 | 0.0102 |
| 132.686 | 34.858 | 142.984 | 0.0055 | 137.338 | 0.0056 |
| 138.4 | 34.99 | 169.848 | 0.0268 | 169.474 | 0.0256 |
| 135.711 | 34.699 | 174.631 | 0.0164 | 173.055 | 0.0156 |
| 134.167 | 34.659 | 155.283 | 0.0062 | 149.496 | 0.0076 |
| 131.357 | 33.538 | 155.018 | 0.011 | 156.591 | 0.0098 |
| 138.468 | 35.444 | 166.861 | 0.0249 | 172.642 | 0.0212 |
| 131.441 | 32.899 | 172.344 | 0.0179 | 174.608 | 0.0168 |
| 132.782 | 33.96 | 144.737 | 0.0158 | 142.038 | 0.0157 |
| 132.386 | 35.034 | 167.418 | 0.0047 | 138.512 | 0.0046 |
| 137.437 | 34.968 | 165.818 | 0.0238 | 167.344 | 0.0237 |
| 132.983 | 32.87 | 152.541 | 0.0413 | 152.891 | 0.0411 |
| 132.967 | 32.756 | 150.881 | 0.0424 | 151.716 | 0.0414 |
| 136.061 | 35.468 | 176.159 | 0.0113 | 171.384 | 0.0102 |
| 132.832 | 32.962 | 155.488 | 0.0388 | 154.089 | 0.0377 |
| 132.532 | 34.07 | 142.344 | 0.0128 | 140.864 | 0.0119 |
| 137.482 | 35.398 | 168.111 | 0.0186 | 168.563 | 0.0184 |
| 134.588 | 35.096 | 159.578 | 0.007 | 153.999 | 0.0069 |
| 133.532 | 33.627 | 151.315 | 0.0288 | 152.054 | 0.0294 |
| 138.101 | 35.045 | 171.559 | 0.0234 | 170.105 | 0.0255 |
| 137.137 | 35.48  | 172.128 | 0.0185 | 170.318 | 0.0164 |
| 136.401 | 34.856 | 173.707 | 0.0188 | 176.462 | 0.0185 |
| 136.07  | 35.321 | 172.771 | 0.0123 | 177.346 | 0.0124 |
| 135.749 | 34.157 | 171.555 | 0.0214 | 173.062 | 0.0204 |
| 132.824 | 34.573 | 136.5   | 0.0081 | 136.921 | 0.0074 |
| 136.603 | 35.229 | 177     | 0.0183 | 176.564 | 0.0184 |
| 138.197 | 34.849 | 167.103 | 0.027  | 166.82  | 0.0262 |
| 135.736 | 34.54  | 171.999 | 0.0194 | 173.331 | 0.0169 |
| 131.328 | 32.685 | −176.904 | 0.0159 | 176.616 | 0.0154 |
| 136.559 | 35.373 | 174.77  | 0.0159 | 177.597 | 0.0158 |
| 135.614 | 35.464 | 166.938 | 0.0083 | 157.565 | 0.0078 |
| 132.793 | 33.172 | 155.083 | 0.0331 | 154.852 | 0.0335 |
| 136.839 | 35.191 | 170.552 | 0.0194 | 172.935 | 0.0199 |
| 131.629 | 32.853 | 171.692 | 0.0192 | 172.275 | 0.02   |
| 137.653 | 34.702 | 164.602 | 0.0279 | 162.043 | 0.0277 |
| 134.322 | 35.101 | 130.527 | 0.0083 | 150.501 | 0.0061 |
| 137.887 | 35.161 | 169.877 | 0.0228 | 171.667 | 0.0219 |
| 134.123 | 33.316 | 154.767 | 0.0452 | 153.072 | 0.0452 |
| 136.56  | 34.548 | 171.172 | 0.0223 | 172.481 | 0.022  |
| 136.017 | 33.742 | 167.332 | 0.0282 | 167.32  | 0.0308 |
| 135.779 | 35.054 | 174.172 | 0.0126 | 172.712 | 0.0128 |
| 136.704 | 35.052 | 172.324 | 0.02   | 173.509 | 0.0203 |
| 138.14  | 34.08  | 169.458 | 0.0497 | 161.4   | 0.0394 |
| 137.61  | 33.88  | 168.303 | 0.0493 | 162.218 | 0.0422 |
| 137.39  | 34.18  | 170.574 | 0.0513 | 173.493 | 0.0516 |
| 137     | 33.67  | 169.653 | 0.0362 | 171.425 | 0.038  |
| 136.67  | 33.43  | 166.896 | 0.0432 | 164.048 | 0.0404 |
| 136.36  | 33.33  | 165.161 | 0.0398 | 168.148 | 0.0394 |
| 135.57  | 33.16  | 159.545 | 0.0466 | 160.163 | 0.0455 |
| 134.94  | 33.35  | 163.591 | 0.034  | 160.607 | 0.0363 |
| 134.81  | 32.87  | 165.677 | 0.0388 | 166.005 | 0.0386 |
| 133.67  | 32.82  | 148.73  | 0.0545 | 149.146 | 0.0544 |
| 134.03  | 32.43  | 149.649 | 0.0481 | 150.022 | 0.0482 |
| 133.22  | 32.37  | 154.515 | 0.0451 | 154.755 | 0.0459 |
| 133.58  | 31.93  | 156.051 | 0.0421 | 153.463 | 0.0406 |

| 132.42 | 32.38 | 145.759 | 0.0375 | 146.819 | 0.0366 |
| 132.49 | 31.97 | 170.491 | 0.02 | 169.559 | 0.0205 |